\documentclass[12pt]{iopart}
\usepackage{iopams}
\expandafter\let\csname equation*\endcsname\relax

\expandafter\let\csname endequation*\endcsname\relax
\usepackage{amssymb,amsmath,amsthm,color}
\usepackage{mathptmx}
\usepackage{amsfonts,bm}
\usepackage{graphicx}
\usepackage[colorlinks=true, linkcolor=red]{hyperref}

\usepackage{tikz}
\usetikzlibrary{matrix}
\usetikzlibrary{positioning}
\usetikzlibrary{intersections,calc}
\newcommand{\myGlobalTransformation}[2]

\def\eq{eq.\ }
\def\eqs{eqs.\ }

\newtheorem*{theorem}{Theorem}

\begin{document}
\title[]{A criterion for the existence of Killing vectors in 3D}

\vspace{1cm}

\author{Boris Kruglikov}
\eads{boris.kruglikov@uit.no}
\address{$^1$\,Department of Mathematics and Statistics, Faculty of Science and Technology, UiT the Arctic University of Norway, Tromso 90-37, Norway}
\address{$^2$\,Department of Mathematics and Natural Sciences, University of Stavanger, 40-36 Stavanger, Norway}
\author{Kentaro Tomoda}
\eads{k-tomoda@sci.osaka-cu.ac.jp}
\address{Advanced Mathematical Institute, Osaka City University, Osaka 558-8585 Japan}

\vspace{1cm}

\begin{abstract}
A three-dimensional Riemannian manifold has locally 6, 4, 3, 2, 1 or none independent Killing vectors.
We present an explicit algorithm for computing dimension of the infinitesimal isometry algebra.
It branches according to the values of curvature invariants. These are relative differential invariants 
computed via curvature, but they are not scalar polynomial Weyl invariants.
We compare our obstructions to the existence of Killing vectors with the known 
existence criteria due to Singer, Kerr and others.
\end{abstract}

\maketitle

\section{Introduction}

The problem of determining the number of Killing tensors (KTs)
of a given Riemannian metric $g_{ab}$ on a manifold $M$ is classical.
Such tensors give first integrals polynomial in momenta and allow integrating the geodesic flow of $g_{ab}$.
The integrals linear in momenta are tantamount to Killing vectors
(KVs) and they generate the isometry algebra of $g_{ab}$.

\begin{figure}[ht]
		\begin{center}
			\begin{tikzpicture}[]
			\matrix (tree) [
			matrix of nodes,
			minimum size=0.4cm,
			column sep=0.4cm,
			row sep=0.4cm,]
			{
				$\blacktriangleright ~ dR = 0$							 & \fbox{3 KVs} \\
				$dR \wedge d[ (\nabla_a R) (\nabla^aR)]= 0$& \fbox{no KV}\\
				$dR \wedge d \Delta R = 0$								& \fbox{1 KV} \\
			};
			\draw[->] (tree-1-1) -- (tree-1-2) node [midway,above] {{\scriptsize yes}};
			\draw[->] (tree-1-1) -- (tree-2-1) node [midway,left] {{\scriptsize no}};
			\draw[->] (tree-2-1) -- (tree-2-2) node [midway,above] {{\scriptsize no}};
			\draw[->] (tree-2-1) -- (tree-3-1) node [midway,left] {{\scriptsize yes}};
			\draw[->] (tree-3-1) -- (tree-3-2) node [midway,above] {{\scriptsize yes}};
			\draw[->] (tree-3-1) -- (tree-2-2) node [midway,above] {{\scriptsize no}};
			\end{tikzpicture}
			\caption{The algorithm for a 2-dimensinal space(-time).
				A triangle symbol $\blacktriangleright$ stands for a root of this algorithm.
				A reverse delta denotes Laplacian, $\Delta \equiv \nabla_a \nabla^a$.
			}
		\label{fig:2-dim}
		\end{center}
	\end{figure}
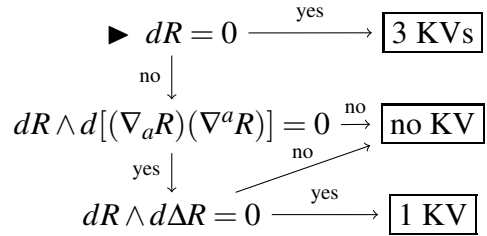

For surfaces G.\,Darboux found a criterion of local existence of KVs that we recall in Figure 
\ref{fig:2-dim} below (the scalar curvature $R$ of $g_{ab}$ can be equivalently changed to the 
Gaussian curvature), see \cite{D} and also \cite{E}.
The corresponding problem for KTs of order 2 is much more involved \cite{HM}.
While a principal approach was sketched in \cite{Sl}, it was only relatively recently
that the final solution was found in \cite{K}, including specification of the
number of KTs of order 2 depending on the curvature invariants of $g_{ab}$.
Criteria for the existence of higher order KTs in dimension 2 and KTs of order 2 
in general dimension are overly complicated; see \cite{HTY} for further discussion.

In this paper, we devise an algorithm for computing the number of local KVs for Riemannian 
manifolds $(M,g_{ab})$ of dimension 3.
The problem of finding dimension of the isometry algebra of $g_{ab}$ is of fundamental importance 
in applications to general relativity. It was addressed before in arbitrary dimension,
so let us recall the status of knowledge.

In \cite{S} I.\,Singer characterised homogeneous spaces locally via the Riemann tensor $R_{abc}{}^d$
and its covariant derivatives.
The problem was later revisited by K.\,Nomizu \cite{N}, F.\,Pr\"ufer-F.\,Tricerri-L.\,Vanhecke \cite{PTV} 
and S.\,Console-C.\,Olmos \cite{CO}.
In the latter two references the homogeneity was given via scalar Weyl invariants of $g_{ab}$.

Recall that scalar Weyl (or polynomial curvature) invariants $I$ are obtained from the covariant derivatives 
of the Riemann tensor $\nabla_{a_1} \cdots \nabla_{a_\sigma} R_{bcd}{}^e$ by tensor products and complete contractions; such invariants are said to have order $\sigma$.
As proved by H.\,Weyl \cite{W} these are the only invariants of $g_{ab}$ polynomial in derivatives 
of the metric components $g_{ij}$.

Using those, R.\,Kerr \cite{Ke} found a criterion for the existence of KVs and expressed
dimension of the isometry algebra in terms of the curvature.
The criterion is as follows. Let $\mathcal{I}_s=\{I:\text{ord}(I)\leq s\}$ be the set of scalar Weyl invariants 
of order $\leq s$.
For a point $o\in M$ denote $\mathcal{O}^s_o=\{x\in M:I(x)=I(o), \forall I\in\mathcal{I}_s\}$,
$\mathcal{R}^s_o=\{A\in SO(T_oM,g): 
A \text{ preserves } \nabla_{a_1} \cdots \nabla_{a_\sigma} R_{bcd}{}^e(o), \sigma \leq s\}$,
and let $m_s=\dim\mathcal{O}^s_o$, $r_s=\dim\mathcal{R}^s_o$.
Then dimension of the algebra of isometries (or the group of motion) is
\begin{align}
d~=~m_s+r_s\:,
\label{eq:dim_iso}
\end{align}
for sufficiently large $s$.

It is possible to show that generally one should take $s=\binom{n+1}2$ in the above formula, $n=\dim M$;
for determining homogeneity it suffices to take $s=\binom{n}2$, cf. \cite{S,CO}.
Also one can modify tensor invariants in the definition of $\mathcal{R}^s_o$ to the differentials of the scalar invariants
from $\mathcal{I}_s$. Let us count the number of scalar invariants involved in the formula \eqref{eq:dim_iso}.

To simplify the evaluation let us relax the condition of polynomial invariants to general differential invariants, i.e.\ arbitrary
scalar (analytic) functions of the metric components $g_{ij}$ and their derivatives invariant under coordinate changes \cite{Th}.
For Riemannian metrics the number of such invariants was computed using the method of Sophus Lie by K.\,Zorawski \cite{Z} for
dimension $n=2$ and C.\,Haskins \cite{Ha} for higher dimensions $n>2$.

Denote the number of functionally independent invariants of order $\sigma=2+s$ (also independent of invariants of order $<\sigma$) by $\delta_\sigma$.
The result for $n=2$ is this: $\delta_0=\delta_1=0$, $\delta_2=\delta_3=1$ and $\delta_\sigma=\sigma-1$ for $\sigma>3$.
Thus with $s=\binom{3}2=3$ we get $\sigma=5$ and the number of involved invariants of order $\leq5$ in $g_{ab}$ is
$\sum_{\sigma\leq5}\delta_\sigma=1+1+3+4=9$.
This definitely exceeds the number 2 of invariants (of orders 4 and 5) in Figure \ref{fig:2-dim},
so even in this simplest case Kerr's criterion is less effective than that by Darboux.

For dimension $n=3$ the situation is even more striking. The count of invariants is as follows:
$\delta_0=\delta_1=0$, $\delta_2=3$ and $\delta_\sigma=\frac32(\sigma-1)(\sigma+2)$ for $\sigma>2$.
Thus with $s=\binom{4}2=6$ we get $\sigma=8$ and the number of independent invariants of order $\leq8$ in $g_{ab}$ is
$\sum_{\sigma\leq8}\delta_\sigma=3+15+27+42+60+81+105=333$.
This is the lower bound for the number of scalar Weyl invariants involved in Kerr's criterion.
Those have never been written down even for the case $n=3$, $s \leq6$.
Though in theory it is possible to find 333 functionally independent among all Weyl invariants, in practice it is rather non-trivial.
One can construct a list of 960 scalar Weyl invariants using the Hamilton-Cayley theorem applied 
to the space of covariant derivatives of the curvature of order $s\leq6$. 
Due to large size of those invariants, to extract 333 functionally independent of them is
a demanding computational task. To use those further to check the number of KVs is a tremendous calculation.

We aim at a more effective criterion to decide the existence and number of KVs.
To this end we devise an algorithm that brings the Killing equations to involution and branches depending on ranks of the equations in
the prolonged system. The PDE system encoding the condition that vector field $K^a$ is a KV for $g_{ab}$ is
\begin{align}
\pounds_K g_{ab} ~=~ 0 \quad \Leftrightarrow \quad \nabla_{(a}K_{b)}=0\:.
\label{eq:Killing}
\end{align}
Here we raise and lower indices with the help of $g_{ab}$, $\nabla_a$ is the Levi-Civita connection of
$g_{ab}$, and $\pounds_K$ is the Lie derivative along $K$.
The first compatibility conditions of this overdetermined system are \cite{HTY}:
\begin{align}
\pounds_K R_{abc}{}^d ~=~ 0\:.
\label{eq:int_cond}
\end{align}
In dimension 3 the Riemann curvature is expressed through the Ricci tensor $R_{ab}$, and we deduce
\begin{align}
&\pounds_K R^{(1)} ~=~ 0\:,
&&\pounds_K R^{(2)} ~=~ 0\:,
&&\pounds_K R^{(3)} ~=~ 0\:,
\end{align}
where $R^{(1)} = R$, $R^{(2)} = R^a{}_bR^b{}_a$, $R^{(3)} = R^a{}_b R^b{}_c R^c{}_a$ are the principal traces of powers of
the Ricci tensor. Thus the matrix equation
\begin{align}
{\boldsymbol R}_a K^a ~=~ 0\:,
\label{eq:matrix_eq}
\end{align}
must be satisfied. Here we define an {\it obstruction $3\times3$ matrix} ${\boldsymbol R}_a$ as
\begin{align}
{\boldsymbol R}_a \equiv
\begin{pmatrix}
\nabla_a R^{(1)} \\
\nabla_a R^{(2)} \\
\nabla_a R^{(3)}
\end{pmatrix}\:.
\label{eq:r_matrix}
\end{align}

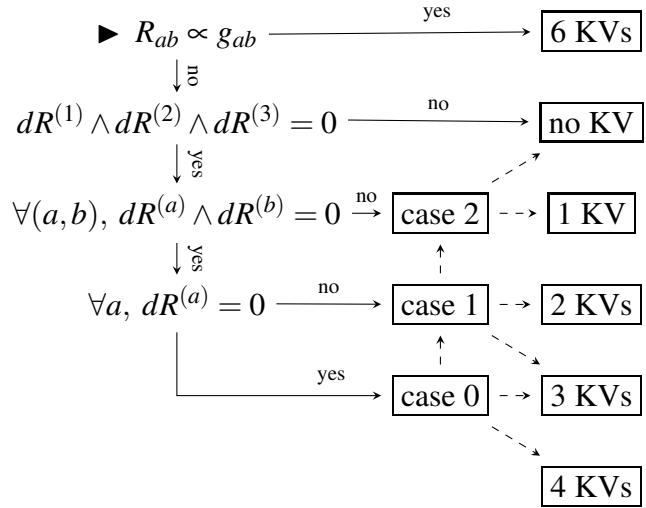
\begin{figure}[h]
	\begin{center}
		\begin{tikzpicture}[>=stealth,sloped]
		\matrix (tree) [
		matrix of nodes,
		minimum size=0.4cm,
		column sep=0.4cm,
		row sep=0.4cm,]
		{
			$\blacktriangleright ~ R_{ab} \propto g_{ab}$ & & \fbox{6 KVs} & \:\\
			$dR^{(1)} \wedge dR^{(2)} \wedge dR^{(3)}= 0$		&  & \fbox{no KV}\\
			$\forall (a,b), ~ dR^{(a)} \wedge dR^{(b)} = 0$		& \fbox{case 2} & \fbox{1 KV}\\
			$\forall a, ~ dR^{(a)} = 0$	& \fbox{case 1} & \fbox{2 KVs}\\
			& \fbox{case 0} & \fbox{3 KVs}\\
			& & \fbox{4 KVs}\\
		};
		\draw[->] (tree-1-1) -- (tree-1-3) node [midway,above] {\hspace{1cm}{\scriptsize yes}};
		\draw[->] (tree-1-1) -- (tree-2-1) node [midway,above] {{\scriptsize no}};
		\draw[->] (tree-2-1) -- (tree-2-3) node [midway,above] {{\scriptsize no}};
		\draw[->] (tree-2-1) -- (tree-3-1) node [midway,above] {{\scriptsize yes}};
		\draw[->] (tree-3-1) -- (tree-3-2) node [midway,above] {{\scriptsize no}};
		\draw[->] (tree-3-1) -- (tree-4-1) node [midway,above] {{\scriptsize yes}};
		\draw[->] (tree-4-1) -- (tree-4-2) node [midway,above] {{\scriptsize no}};
		\draw[->] (tree-4-1) |- (tree-5-2) node [midway,above] {\hspace{4cm} {\scriptsize yes}};
		\draw[->,dashed] (tree-3-2) -- (tree-3-3);
		\draw[->,dashed] (tree-3-2) -- (tree-2-3);
		\draw[->,dashed] (tree-4-2) -- (tree-3-2);
		\draw[->,dashed] (tree-4-2) -- (tree-4-3);
		\draw[->,dashed] (tree-4-2) -- (tree-5-3);
		\draw[->,dashed] (tree-5-2) -- (tree-4-2);
		\draw[->,dashed] (tree-5-2) -- (tree-5-3);
		\draw[->,dashed] (tree-5-2) -- (tree-6-3);
		\end{tikzpicture}
		\caption{Main branching of the algorithm to determine the number of KVs for a 3-dimensinal space.
			Dashed lines include complicated processes which are shown in Figures \ref{fig:case2}--\ref{fig:case0}.
		}
	\label{fig:main}
	\end{center}
\end{figure}

The matrix equation \eqref{eq:matrix_eq} yields an immediate consequence:
Any KV must be in the kernel of ${\boldsymbol R}_a$.
Hence, the determinant of ${\boldsymbol R}_a$
\begin{align}\label{boldR}
\det {\boldsymbol R}_a
~=~& dR^{(1)} \wedge dR^{(2)} \wedge dR^{(3)}\:,
\end{align}
has to vanish. This is a relative differential invariant.
Otherwise $\ker{\boldsymbol R}_a=0$ and consequently there is no KV.

It is clear that the orbit dimension of the motion group action is bounded as
$m=m_s \leq\dim\ker{\boldsymbol R}_a=3-\mathrm{rank} {\boldsymbol R}_a$. 
Since $r=r_s \leq\binom{m}2$,
dimension of this group (= number of KVs) is $d=m+r\leq\binom{m+1}2$, 
and this is bounded via the rank of \eqref{eq:r_matrix}.

We consider cases $2,1,0$, called so according to $\mathrm{rank} {\boldsymbol R}_a =2,1,0$
(respectively, $\dim\ker{\boldsymbol R}_a =1,2,3$).
We examine branching governing the number of KVs depending on the values of scalar curvature 
invariants in all cases.
We summarise our result so:
 \begin{theorem}
Let $(M, g_{ab})$ be a 3-dimensional Riemannian manifold.
The dimension of the isometry algebra can be computed via differential invariants of
the algorithm outlined in Figure \ref{fig:main}.
This algorithm includes further branching given in Figures \ref{fig:case2}--\ref{fig:case0}.
 \end{theorem}
Notice that the differential invariants used in our theorem are rational in derivatives of the metric components $g_{ij}$,
and so they are not necessarily Weyl invariants.
However since we use only differential relations, these can be assumed relative scalar polynomial invariants.

The rest of the paper is composed as follows.
In Sections \ref{sec:case2}--\ref{sec:case0},
we correspondingly give the formulation and proof for the algorithm.
In Section \ref{sec:exam} we present two examples of its application.
In Section \ref{sec:conc} we close this paper with a comment on relations of our method to that of Cartan.
Two appendices 
contain technical formulae.

\section{Analysis of case 2}\label{sec:case2}
In this case, it follows from the rank-nullity theorem that
$\dim \ker {\boldsymbol R}_a = 1$.
So if we have an annihilator of ${\boldsymbol R}_a$, any KV can be written by
\begin{align}
K^a = \omega \:U^a\:,
\label{eq:KV_ansatz_case2}
\end{align}
$\omega$ and $U^a$ are respectively an unknown function and the annihilator.
We take the annihilator as
\begin{align}
U^a ~\equiv~ U\:\epsilon^{abc}
(\nabla_b R^{(1)}) (\nabla_c R^{(2)}) \:,
\label{eq:annihilator_case2}
\end{align}
where the normalisation factor $U$ is determined by
\begin{align}
U^{-2} 
~&=~2(\nabla_{[a} R^{(1)})(\nabla_{b]} R^{(2)}) (\nabla^{[a} R^{(1)}) (\nabla^{b]} R^{(2)})
\:,
\end{align}
so as to satisfy $U_a U^a = 1$.
If $U^a$ vanishes identically,
two scalars $(R^{(1)}, R^{(2)})$ in the definition \eqref{eq:annihilator_case2}
must be replaced by $(R^{(2)}, R^{(3)})$ or $(R^{(3)}, R^{(1)})$.

Using the concrete form \eqref{eq:KV_ansatz_case2},
we write out the components of the Killing equation \eqref{eq:Killing}.
To this end, we introduce the projection tensor onto the hyperplanes orthogonal to $U^a$ as
\begin{align}
q_{ab}(U) ~\equiv~ g_{ab} -U_a U_b\:,
\label{eq:projection}
\end{align}
that is endowed with a projection property and an orthogonality
\begin{align}
&q_{ac} q^c{}_b ~=~ q_{ab}\:,
&&q_{ab}\:U^b ~=~ 0\:.
\end{align}
The $UU$,  $Uq$ and $q q$-parts
of the Killing equation \eqref{eq:Killing}
have respectively $1$, $2$ and $3$ components as follows.
\begin{subequations}
\begin{align}
0~=~ &U^a U^b \nabla_{(a} K_{b)} ~=~ \pounds_U \omega\:,
\label{eq:Killing_UU_case2}
\\
0~=~ &U^a q^b{}_c \nabla_{(a} K_{b)} ~=~
\tfrac{1}{2}(\nabla_c \omega - (\pounds_U \omega) U_c + \Omega_c \omega)\:,\\
0~=~ &q^a{}_c q^b{}_d \nabla_{(a} K_{b)}
~=~\omega \kappa_{cd}\:,
\end{align}
\end{subequations}
where $\pounds_U$ is the Lie derivative along $U^a$, $\Omega_a$ and $\kappa_{ab}$ are defined as
\begin{align}
&\Omega_a(U) ~\equiv~ U^b \nabla_b U_a\:,
&& \kappa_{ab}(U) ~\equiv~
q^c{}_a q^d{}_b \nabla_{(c} U_{d)}\:.
\label{eq:obsts_case2}
\end{align}
It can be concluded that
the Killing equation \eqref{eq:Killing} can be rewritten as
\footnote[1]{Note that the second equation implies \eq \eqref{eq:Killing_UU_case2}.}
\begin{align}
&\kappa_{ab} ~=~0\:,
&&\nabla_a \omega ~=~ -\Omega_a \omega\:.
\label{eq:Killing_case2}
\end{align}
The integrability condition for \eq \eqref{eq:Killing_case2} is given by
\begin{align}
\nabla_{[a} \Omega_{b]} ~=~ 0\:.
\end{align}

\begin{figure}[t]
	\begin{center}
		\begin{tikzpicture}[>=stealth,sloped]
		\matrix (tree) [
		matrix of nodes,
		minimum size=0.4cm,
		column sep=0.4cm,
		row sep=0.4cm,]
		{
			$\blacktriangleright \ \kappa_{ab} = \nabla_{[a} \Omega_{b]} = 0$ & \fbox{1 KV} \\
			& \fbox{no KV} \\
		};
		\draw[->] (tree-1-1) -- (tree-1-2) node [midway,above] {{\scriptsize yes}};
		\draw[->] (tree-1-1) |- (tree-2-2) node [midway,above] {\hspace{1.5cm}{\scriptsize \qquad no}};
		\end{tikzpicture}
		\caption{Our sub-algorithm for case 2,
			see \eq \eqref{eq:obsts_case2} for notations.}
	\label{fig:case2}
	\end{center}
\end{figure}
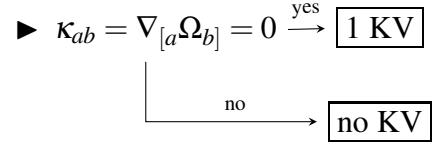

If the annihilator $U^a$ passes the two tests,
\begin{align}
&\kappa_{ab} ~=~0\:,
&&\nabla_{[a} \Omega_{b]} ~=~ 0\:,
\label{eq:tests_case2}
\end{align}
then there are no extra conditions
that must be satisfied, thereby allowing us to confirm that one KV exists.
The results obtained here are summarised in Figure \ref{fig:case2}

Observe that our tests \eqref{eq:tests_case2} do not depend on
the explicit form of $U^a$, \eq \eqref{eq:annihilator_case2}.
Therefore if we can write KVs in the form of \eq \eqref{eq:KV_ansatz_case2},
our analysis here will be recyclable.

\section{Analysis of case 1}\label{sec:case1}
Again by the rank-nullity theorem, $\dim \ker {\boldsymbol R}_a = 2$.
Then KVs take the form
\begin{align}
K^a =  \omega_N\:N^a +  \omega_B\:B^a\:,
\label{eq:KV_ansatz_case1}
\end{align}
where $(N^a, B^a)$ are two annihilators of ${\boldsymbol R}_a$ and
$(\omega_N, \omega_B)$ are two unknown functions.
We assume that $(N^a, B^a)$ are unit vector fields satisfying an orthogonality $N^a B_a = 0$.
In order to provide a versatile algorithm,
we do not make any more assumptions.

By combining $(N^a, B^a)$ and a unit vector field $T^a$ defined as
\begin{align}
T^a ~\equiv~ \frac{\nabla^a R^{(i)}}{\sqrt{(\nabla_b R^{(i)})(\nabla^b R^{(i)})}}\:,
\label{eq:T_case1}
\end{align}
for the non-zero $\nabla_aR^{(i)}~(i = 1, 2$ or $3)$, we obtain
a nonholonomic orthonormal basis $(T^a, N^a, B^a)$,
\begin{align}
\delta^a{}_b ~=~ T^a T_b + N^a N_b + B^a B_b\:.
\label{eq:frame_case1}
\end{align}
By using \eqs \eqref{eq:KV_ansatz_case1} and \eqref{eq:frame_case1},
the $TT$-part of the Killing equation can formally be written by
\begin{align}
0&~=~ (\omega_N N^a+\omega_B B^a)~T^b\nabla_b T_a\:,
\label{eq:KillingTT_case1}
\end{align}
which gives a test for $R^{(i)}$.
If there are two linearly independent KVs, then \eq \eqref{eq:KillingTT_case1} implies $T^b \nabla_b T_a = 0$.
Depending on whether the gradient of $R^{(i)}$
satisfies the geodesic equation
\begin{align}
(\nabla^b R^{(i)}) \nabla_b \nabla_a R^{(i)} ~\propto~ \nabla_a R^{(i)}\:,
\end{align}
our analysis branches off.

\subsection{Branch where $\nabla_a R^{(i)}$ is not a geodesic}
In this branch $T^a$ and
its acceleration $T^b \nabla_b T^a$ are linearly independent.
It is therefore possible to define the Frenet--Serret frame as
\begin{align}
&T^a ~\equiv~ \frac{\nabla^a R^{(i)}}{\sqrt{(\nabla_b R^{(i)})(\nabla^b R^{(i)})}}\:,
&&N^a ~\equiv~
\frac{T^b \nabla_b T^a}{\sqrt{(T^c \nabla_c T^e)(T^d \nabla_d T_e) }}\:,
&&B^a ~\equiv~ \epsilon^{abc} T_b N_c\:.
\end{align}
This frame obeys the so-called Frenet-–Serret formulae
\begin{align}
T^b \nabla_b
\begin{pmatrix}
T^a \\
N^a \\
B^a
\end{pmatrix}
~=~
\begin{pmatrix}
0 	& \kappa_T & 0 \\
-\kappa_T& 0 & \tau_T \\
0	& -\tau_T & 0
\end{pmatrix}
\begin{pmatrix}
T^a \\
N^a \\
B^a
\end{pmatrix}\:,
\end{align}
where
\begin{align}
&\kappa_T ~\equiv~ N^a T^b \nabla_b T_a\:,
&&\tau_T ~\equiv~ B^aT^b \nabla_b N_a\:,
\end{align}
are respectively the geodesic curvature
and torsion of an integral curve of $T^a$.

Now, the $TT$-part of the Killing equation \eqref{eq:KillingTT_case1} reads
\begin{align}
\kappa_T\:\omega_N ~=~ 0\:.
\end{align}
Since $\kappa_T = 0$ contradicts $T^b \nabla_b T^a \neq 0$,
thus $\omega_N$ must be zero.
As KVs take the form $K^a = \omega_B B^a$,
our analysis reduces to that of case 2 with
the identification of $B^a \rightarrow U^a$.
Hence, $\Omega^a (B)$ and $\kappa_{ab} (B)$ defined in \eq \eqref{eq:obsts_case2}
give the tests for $B^a$.
In this branch, there is at most one KV.

\subsection{Branch where $\nabla_a R^{(i)}$ is a geodesic}

Although the following analysis does not depend on the choice of the orthonormal basis $(T^a, N^a, B^a)$,
we shall comment on it at any rate.
For Riemannian metrics, there are two natural bases depending on the property of $T^a$.
If $T^a$ is an eigenvector of the Ricci tensor $R^{a}{}_b$, we can take the orthonormal basis
as the eigensystem of $R^{a}{}_b$.
Otherwise $(N^a, B^a)$ can be taken to be
\begin{align}
&N^a ~\equiv~ N\:\epsilon^{abc} T_b (R^d{}_cT_d)\:,
&&B^a ~\equiv~ \epsilon^{abc} T_b N_c\:,
\end{align}
since $T^a$ and $R^a{}_b T^b$ are linearly independent.
Here $N$ is the normalisation factor.

In this branch, the $TT$-part of the Killing equation \eqref{eq:KillingTT_case1}
is identically satisfied.
To write out the remaining parts,
we introduce the Ricci rotation coefficients as
\begin{subequations}
\begin{align}
T^b \nabla_b
\begin{pmatrix}
T^a \\
N^a \\
B^a
\end{pmatrix}
&~=~
\begin{pmatrix}
0 	& 0 & 0 \\
0	& 0 & \tau_T \\
0	& -\tau_T & 0
\end{pmatrix}
\begin{pmatrix}
T^a \\
N^a \\
B^a
\end{pmatrix}\:,\\
N^b \nabla_b
\begin{pmatrix}
T^a \\
N^a \\
B^a
\end{pmatrix}
&~=~
\begin{pmatrix}
0& -\kappa_N & \tau_N \\
\kappa_N	& 0 & \eta_N \\
-\tau_N	& -\eta_N & 0
\end{pmatrix}
\begin{pmatrix}
T^a \\
N^a \\
B^a
\end{pmatrix}\:, \\
B^b \nabla_b
\begin{pmatrix}
T^a \\
N^a \\
B^a
\end{pmatrix}
&~=~
\begin{pmatrix}
0& \tau_B & -\kappa_B \\
-\tau_B & 0 & -\eta_B \\
\kappa_B& \eta_B & 0
\end{pmatrix}
\begin{pmatrix}
T^a \\
N^a \\
B^a
\end{pmatrix}\:,
\end{align}
\end{subequations}
where
\begin{align}
&\kappa_N ~\equiv~ T^aN^b \nabla_b N_a\:,
&&\eta_N ~\equiv~ B^aN^b \nabla_b N_a\:,
&&\tau_N ~\equiv B^aN^b \nabla_b T_a\:,\notag\\
&\kappa_B ~\equiv~ T^aB^b \nabla_b B_a\:,
&&\eta_B ~\equiv~ N^aB^b \nabla_b B_a\:,
&&\tau_B ~\equiv N^aB^b \nabla_b T_a\:.
\label{eq:ricci_coes_case1}
\end{align}
$\kappa_N (\kappa_B)$, $\eta_N (\eta_B)$ and $\tau_N (\tau_B)$
are respectively the geodesic, normal curvature and relative torsion
of an integral curve of $N^a (B^a)$.
Since the derivatives of the Ricci rotation coefficients are not independent,
we collect their relations in \ref{app:curv_rels_case1}.

Using the Ricci rotation coefficients, the remaining parts of
the Killing equation read
\begin{subequations}
	\begin{align}
	\pounds_T\omega_N ~=~& -\kappa_N \omega_N +(\tau_T + \tau_N) \omega_B\:,\\
	\pounds_N\omega_N ~=~& \eta_N \omega_B \:,\\
	\pounds_B\omega_N ~=~&  -\eta_B \omega_B -\eta_N \omega_N -\bar{\omega}\:,\\
	\pounds_T\omega_B ~=~& - (\tau_T - \tau_B) \omega_N - \kappa_B \omega_B\:,\\
	\pounds_N\omega_B ~=~& \bar{\omega} \:,
	\label{eq:bar_omega_case1}
	\\
	\pounds_B\omega_B ~=~& \eta_B \omega_N\:,
	\end{align}
	\label{eq:Killing_case1_sub}
\end{subequations}
where \eq \eqref{eq:bar_omega_case1} defines a new variable $\bar{\omega}$.
Clearly, the above equations are not closed with respect to unknown scalars $(\omega_N, \omega_B, \bar{\omega})$.
We thus need the information of the second order derivatives of $(\omega_N, \omega_B)$.
From the identities $\nabla_{[a} \nabla_{b]} \omega_N = \nabla_{[a} \nabla_{b]} \omega_B = 0$,
we obtain the constraint equations
\begin{subequations}
\begin{align}
0~=~& 2(\tau_B - \tau_N) \bar{\omega}
+\Bigl[ 2\eta_N (\tau_B-\tau_N) -\pounds_B(\tau_B - \tau_N)
+\pounds_N (\kappa_B+\kappa_N) \Bigr]\omega_N \notag \\
&+\Bigl[\pounds_B(\kappa_B + \kappa_N) + \pounds_N (\tau_B-\tau_N)\Bigr] \omega_B\:,
\label{eq:int_eqs1_case1}\\
0~=~& 2(\tau_B + \tau_N) \bar{\omega}
+\Bigl[\pounds_N (\kappa_B-\kappa_N) -\pounds_B(\tau_B - \tau_N) \Bigr]\omega_N \notag\\
&+\Bigl[2 \eta_B (\tau_B+\tau_N) + \pounds_B(\kappa_B-\kappa_N)
-\pounds_N(\tau_B-\tau_N) \Bigr] \omega_B\:,
\label{eq:int_eqs2_case1}\\
0~=~& (\kappa_B-\kappa_N)\bar{\omega}
-(\pounds_N \tau_B) \omega_N
+\Bigl[\eta_B (\kappa_B-\kappa_N) -\pounds_B \tau_N\Bigr] \omega_B\:.
\label{eq:int_eqs3_case1}
\end{align}
\end{subequations}
Notice that the remaining parts of $\nabla_{[a} \nabla_{b]} \omega_N = \nabla_{[a} \nabla_{b]} \omega_B = 0$
constitute the equations of evolution of $\bar{\omega}$.
However, if the constraints \eqref{eq:int_eqs1_case1}--\eqref{eq:int_eqs3_case1} are not satisfied,
we do not need to take such equations.

\subsubsection{Sub-branch where $\tau_N = \tau_B = \kappa_N - \kappa_B = 0$}

In this sub-branch,
unknown functions $( \omega_N,\omega_B,\bar{\omega})$ are
free from the constraints \eqref{eq:int_eqs1_case1}--\eqref{eq:int_eqs3_case1}.
Thus, we look at the remaining parts of the identities $\nabla_{[a} \nabla_{b]} \omega_N = \nabla_{[a} \nabla_{b]} \omega_B = 0$
and obtain
\begin{subequations}
\begin{align}
\pounds_T \bar{\omega} ~=~&
(\eta_N \kappa_N + \pounds_B \kappa_N - \pounds_T \eta_N) \omega_N
-(\eta_N \tau_T + \pounds_N \kappa_N) \omega_B \:, \\
\pounds_N \bar{\omega} ~=~&
-(\pounds_N \eta_N) \omega_N
+(\eta_B^2 - \pounds_B \eta_N - \pounds_N \eta_B) \omega_B \:, \\
\pounds_B \bar{\omega} ~=~&
(\pounds_N \eta_B-\eta_B^2) \omega_N
+\eta_N \eta_B\omega_B + \eta_N \bar{\omega}\:.
\end{align}
\end{subequations}
Therefore, the equations of evolution of $(\omega_N, \omega_B, \bar{\omega})$
take the form
\begin{align}
&\nabla_a {\boldsymbol \omega} ~=~{\boldsymbol \Omega}_a^{(1)} {\boldsymbol \omega}\:,
&&{\boldsymbol \omega} ~\equiv~
\begin{pmatrix}
\omega_N \\
\omega_B \\
\bar{\omega}
\end{pmatrix}\:,
\label{eq:Killing_case1_submaximal}
\end{align}
where
	\begin{align}
	{\boldsymbol \Omega}_a^{(1)} ~\equiv~&
	T_a
	\begin{pmatrix}
	-\kappa_N & \tau_T & 0\\
	-\tau_T & -\kappa_N & 0 \\
	\eta_N \kappa_N + \pounds_B \kappa_N -\pounds_T \eta_N
	&-\eta_N \tau_T - \pounds_N \kappa_N& 0
	\end{pmatrix} \\
	\notag
	&+N_a
	\begin{pmatrix}
	0 & \eta_N & 0\\
	0 & 0 & 1 \\
	-\mathcal{L}_N \eta_N
	&\eta_B^2 -\pounds_B \eta_N - \pounds_N \eta_B
	& 0
	\end{pmatrix}
	+B_a
	\begin{pmatrix}
	-\eta_N & -\eta_B & -1\\
	\eta_B & 0 & 0 \\
	\pounds_N \eta_B -\eta_B^2&
	\eta_N \eta_B&
	\eta_N
	\end{pmatrix}
	\:.
	\end{align}
The integrability condition for \eq \eqref{eq:Killing_case1_submaximal} reads
\begin{align}
\bigl( \nabla_{[a} {\boldsymbol \Omega}_{b]}^{(1)}
- {\boldsymbol \Omega}^{(1)}_{[a} {\boldsymbol \Omega}_{b]}^{(1)}
\bigr){\boldsymbol \omega} ~=~ 0\:,
\end{align}
or in component form
\begin{align}
{\boldsymbol R}^{(1)}_{\mathrm{cs. 1}} {\boldsymbol \omega} ~=~ 0\:,
\label{eq:IntCond_case1_submaximal_1}
\end{align}
where
\begin{align}
{\boldsymbol R}^{(1)}_{\mathrm{cs. 1}}~\equiv~
\begin{pmatrix}
\pounds_N \kappa_N & \pounds_B \kappa_N & 0\\
\pounds_N \lambda_N & \pounds_B \lambda_N & 0
\end{pmatrix}\:,
\label{eq:1st_obst_case1}
\end{align}
with $\lambda_N ~\equiv~ R_{ab} N^a N^b$.
We call ${\boldsymbol R}^{(1)}_{\mathrm{cs. 1}}$ the {\it first obstruction matrix of case 1}
whose rank governs the number of KVs.
If $\mathrm{rank} {\boldsymbol R}^{(1)}_{\mathrm{cs. 1}} = 0$, 3 KVs exist.
If $\mathrm{rank} {\boldsymbol R}^{(1)}_{\mathrm{cs. 1}} = 2$, there is no KV.
If $\mathrm{rank} {\boldsymbol R}^{(1)}_{\mathrm{cs. 1}} = 1$,
our analysis reduces to that of case 2 with appropriate identifications of $U^a$.
\footnote{
		For instance, if $\pounds_N \kappa_N \neq 0$,
		from \eq \eqref{eq:IntCond_case1_submaximal_1} we can write
		$\omega_N = -\frac{\pounds_B \kappa_N}{\pounds_N \kappa_N} \omega_B$
		and then KVs take the form
		$K^a ~=~ \omega_B \left(B^a - \frac{\pounds_B \kappa_N}{\pounds_N \kappa_N} N^a\right)$.
		Thus, under the identification
		$\alpha\left(B^a - \frac{\pounds_B \kappa_N}{\pounds_N \kappa_N} N^a\right)
		\rightarrow U^a$,where $\alpha$ is the normalisation factor, our analysis reduces to that of case 2.}

\subsubsection{Sub-branch where $\tau_N = \tau_B = 0$ but $\kappa_N \neq \kappa_B$}
In this sub-branch, from \eq \eqref{eq:int_eqs3_case1}
the function $\bar{\omega}$ takes the form
\begin{align}
\bar{\omega}~=~ -\eta_B \omega_B\:.
\end{align}
Substituting this form into \eqs \eqref{eq:Killing_case1_sub},
we obtain
\begin{align}
\nabla_a
{\boldsymbol \omega}
~=~
{\boldsymbol \Omega}_a^{(2)}
{\boldsymbol \omega}
\:,
&&
{\boldsymbol \omega}~\equiv~
\begin{pmatrix}
\omega_N \\
\omega_B
\end{pmatrix}\:,
\label{eq:Killing_case1_sub2}
\end{align}
where
\begin{align}
{\boldsymbol \Omega}_a^{(2)}
~\equiv~
T_a
\begin{pmatrix}
-\kappa_N & \tau_T\\
-\tau_T & -\kappa_B
\end{pmatrix}
+N_a
\begin{pmatrix}
0 & \eta_N\\
0 & -\eta_B
\end{pmatrix}
+B_a
\begin{pmatrix}
-\eta_N & 0\\
\eta_B & 0
\end{pmatrix}
\:.
\end{align}
Its integrability condition
$( \nabla_{[a} {\boldsymbol \Omega}^{(2)}_{b]}- {\boldsymbol \Omega}^{(2)}_{[a} {\boldsymbol \Omega}^{(2)}_{b]}){\boldsymbol \omega} = 0$
leads to
${\boldsymbol R}^{(2)}_{\mathrm{cs. 1}} {\boldsymbol \omega} = 0$
where
\begin{align}
{\boldsymbol R}^{(2)}_{\mathrm{cs. 1}} ~\equiv~
\begin{pmatrix}
\pounds_N \kappa_N & \pounds_B \kappa_N\\
\pounds_N \kappa_B & \pounds_B \kappa_B\\
\pounds_N \tau_T & \pounds_B \tau_T\\
\pounds_N \eta_N & \pounds_B \eta_N\\
\pounds_N \eta_B & \pounds_B \eta_B
\end{pmatrix}
\:.
\label{eq:2nd_obst_case1}
\end{align}
We call ${\boldsymbol R}^{(2)}_{\mathrm{cs. 1}}$ the {\it second obstruction matrix of case 1}.
In a way analogous to ${\boldsymbol R}^{(1)}_{\mathrm{cs. 1}}$,
$\mathrm{rank} {\boldsymbol R}^{(2)}_{\mathrm{cs. 1}}$ reveals the number of KVs.
If $\mathrm{rank} {\boldsymbol R}^{(2)}_{\mathrm{cs. 1}} = 0$, two KVs exist.
If $\mathrm{rank} {\boldsymbol R}^{(2)}_{\mathrm{cs. 1}}= 2$, there is no KV.
Otherwise, when $\mathrm{rank} {\boldsymbol R}^{(2)}_{\mathrm{cs. 1}} = 1$,
our analysis reduces to that of case 2 with
an appropriate identification of $U^a$.

\subsubsection{Sub-branch where $\tau_N = \tau_B \neq 0$}

In this sub-branch, it follows from \eq \eqref{eq:int_eqs2_case1} that
\begin{align}
\bar{\omega}~=~ -\eta_B \omega_B
+\frac{1}{4\tau_N} \Bigl[
\omega_N \pounds_N (\kappa_N -\kappa_B)
+ \omega_B \pounds_B (\kappa_N -\kappa_B)
\Bigr]\:.
\end{align}
By using this, we rewrite \eqs \eqref{eq:Killing_case1_sub} as
\begin{align}
&\nabla_a
{\boldsymbol \omega}
~=~
{\boldsymbol \Omega}_a^{(3)} {\boldsymbol \omega}\:,
&&
{\boldsymbol \omega}~\equiv~
\begin{pmatrix}
\omega_N \\
\omega_B
\end{pmatrix}\:,
\label{eq:Killing_case1_sub3}
\end{align}
where
\begin{align}
{\boldsymbol \Omega}_a^{(3)}
~\equiv~&
T_a
\begin{pmatrix}
-\kappa_N & \tau_T + \tau_N\\
-\tau_T+\tau_N & -\kappa_B
\end{pmatrix}
+N_a
\begin{pmatrix}
0 & \eta_N\\
\tfrac{1}{4\tau_N} \pounds_N (\kappa_N -\kappa_B)&
-\eta_B + \tfrac{1}{4\tau_N} \pounds_B(\kappa_N-\kappa_B)
\end{pmatrix}
\notag \\
&+B_a
\begin{pmatrix}
-\eta_N -\tfrac{1}{4\tau_N} \pounds_N (\kappa_N - \kappa_B)
& -\eta_B -\tfrac{1}{4\tau_N} \pounds_B (\kappa_N - \kappa_B)\\
\eta_B & 0
\end{pmatrix}
\:.
\end{align}
Using the shorthand notation for the Ricci rotation coefficients,
\begin{align}
&\kappa_\Delta~\equiv~\kappa_B-\kappa_N\:,
&&\kappa_\Sigma ~\equiv~ \kappa_B+\kappa_N\:,
\label{eq:shorthand_case1}
\end{align}
the integrability condition for \eq \eqref{eq:Killing_case1_sub3} can be written as follows.
\begin{subequations}
\begin{align}
0~=~&(\pounds_N \kappa_\Sigma) \omega_N+(\pounds_B \kappa_\Sigma) \omega_B\:,\\
0~=~&\left[\pounds_N (\kappa_\Delta^2+4\tau_N^2) \right] \omega_N
+\left[ \pounds_B (\kappa_\Delta^2+4\tau_N^2) \right] \omega_B \:,\\
0~=~&\left[ \pounds_N \pounds_T \kappa_\Delta
-4\tau_N \pounds_N \tau_T-\frac{(\pounds_T\tau_N) (\pounds_N \kappa_\Delta)}{\tau_N}\right]
\omega_N\notag\\
&+\left[\pounds_B \pounds_T \kappa_\Delta
-4\tau_N \pounds_B \tau_T-\frac{(\pounds_T\tau_N) (\pounds_B \kappa_\Delta)}{\tau_N}\right] \omega_B\:,\\
0~=~&\left[
\pounds_N \pounds_N \kappa_\Delta
-4\tau_N \pounds_N \eta_N
-\eta_B \pounds_N \kappa_\Delta
-\frac{\pounds_N \kappa_\Delta}{4\tau_N} (4 \pounds_N \tau_N + \pounds_B \kappa_\Delta)
\right] \omega_N\notag\\
&+\left[
\pounds_B \pounds_N \kappa_\Delta
-4\tau_N \pounds_B \eta_N
-\eta_B \pounds_B \kappa_\Delta
-\frac{\pounds_B \kappa_\Delta}{4\tau_N} (4 \pounds_N \tau_N + \pounds_B \kappa_\Delta)
\right] \omega_B\:,\\
0~=~&\left[
\pounds_N \pounds_B \kappa_\Delta
+4\tau_N \pounds_N \eta_B
-\eta_N \pounds_N \kappa_\Delta
-\frac{\pounds_N \kappa_\Delta}{4\tau_N} (4 \pounds_B \tau_N - \pounds_N \kappa_\Delta)
\right] \omega_N\notag\\
&+\left[
\pounds_B \pounds_B \kappa_\Delta
+4\tau_N \pounds_B \eta_B
-\eta_N \pounds_B \kappa_\Delta
-\frac{\pounds_B \kappa_\Delta}{4\tau_N} (4 \pounds_B \tau_N - \pounds_N \kappa_\Delta)
\right] \omega_B\:,
\end{align}
\label{eq:3rd_obst_case1}
\end{subequations}
Rewriting \eqs \eqref{eq:3rd_obst_case1} as
${\boldsymbol R}^{(3)}_{\mathrm{cs. 1}} {\boldsymbol \omega} = 0$,
the rank of the {\it third obstruction matrix of case 1} ${\boldsymbol R}^{(3)}_{\mathrm{cs. 1}}$
governs the number of KVs in a way analogous to
that of ${\boldsymbol R}^{(1)}_{\mathrm{cs. 1}}$ and ${\boldsymbol R}^{(2)}_{\mathrm{cs. 1}}$.

\subsubsection{Sub-branch where $\tau_N \neq \tau_B$}

As similar to the previous sub-branch,
we can put $\bar{\omega}$ into $(\omega_N, \omega_B)$
by using \eq \eqref{eq:int_eqs1_case1},
\begin{align}
\bar{\omega}~=~
\frac{1}{2\tau_\Delta}\left[
\pounds_B \tau_\Delta
+\tau_\Delta (\eta_\Delta-\eta_\Sigma)
-\pounds_N \kappa_\Sigma
\right]\omega_N
-\frac{1}{2\tau_\Delta}\left[
\pounds_N \tau_\Delta
+\pounds_B \kappa_\Sigma \right]\omega_B
\:,
\end{align}
where we have used the shorthand notations \eqref{eq:shorthand_case1} and
\begin{align}
&\tau_\Delta~\equiv~\tau_B-\tau_N\:,
&&\tau_\Sigma ~\equiv~ \tau_B+\tau_N\:,
&&\eta_\Delta~\equiv~\eta_B-\eta_N\:,
&&\eta_\Sigma ~\equiv~ \eta_B+\eta_N\:.
\end{align}
Thus, we rewrite \eqs \eqref{eq:Killing_case1_sub} as
\begin{align}
&\nabla_a
{\boldsymbol \omega}
~=~
{\boldsymbol \Omega}_a^{(4)} {\boldsymbol \omega}\:,
&&
{\boldsymbol \omega}~\equiv~
\begin{pmatrix}
\omega_N \\
\omega_B
\end{pmatrix}\:,
\label{eq:Killing_case1_sub4}
\end{align}
where
\begin{align}
{\boldsymbol \Omega}_a^{(4)}
~\equiv~&
T_a
\begin{pmatrix}
\tfrac{\kappa_\Delta-\kappa_\Sigma}{2} & \tau_T -\tfrac{\tau_\Delta-\tau_\Sigma}{2}\\
-\tau_T+\tfrac{\tau_\Delta+\tau_\Sigma}{2} &-\tfrac{\kappa_\Delta+\kappa_\Sigma}{2}
\end{pmatrix}
+N_a
\begin{pmatrix}
0 & \tfrac{\eta_\Sigma-\eta_\Delta}{2}\\
\tfrac{\pounds_B \tau_\Sigma -\pounds_N \kappa_\Sigma +\tau_\Delta (\eta_\Delta-\eta_\Sigma)}{2\tau_\Delta}&
-\tfrac{\pounds_B \kappa_\Sigma + \pounds_N \tau_\Delta}{2\tau_\Delta}
\end{pmatrix}
\notag \\
&+B_a
\begin{pmatrix}
\tfrac{\pounds_N \kappa_\Sigma - \pounds_B \tau_\Delta}{2\tau_\Delta}
& \tfrac{\pounds_B \kappa_\Sigma + \pounds_N \tau_\Delta -\tau_\Delta(\eta_\Delta + \eta_\Sigma)}{2\tau_\Delta}\\
\tfrac{\eta_\Delta + \eta_\Sigma}{2} & 0
\end{pmatrix}\:.
\end{align}
Its integrability condition leads to tolerably complicated relations
 {\footnotesize
\begin{subequations}
\begin{align}
0~=~& \Bigl[ \tau_\Delta \left(\pounds_B \tau_\Delta - \pounds_N \kappa_\Delta\right)
+\tau_\Sigma \left( \tau_\Delta (\eta_\Sigma -\eta_\Delta) - \pounds_B \tau_\Delta + \pounds_N \kappa_\Sigma  \right)\Bigr]\omega_N \notag \\
&+\Bigl[
\tau_\Delta\left( \pounds_N \tau_\Delta - \pounds_B \kappa_\Delta \right)
+\tau_\Sigma \left( -\tau_\Delta (\eta_\Delta + \eta_\Sigma) + \pounds_B \kappa_\Sigma + \pounds_N \tau_\Delta\right)
\Bigr]\omega_B\:,\\
0~=~& \Bigl[
\kappa_\Delta \left(\pounds_B \tau_\Delta - \pounds_N \kappa_\Sigma\right)
+\tau_\Delta \left( \kappa_\Delta (\eta_\Delta -\eta_\Sigma)-\pounds_N (\tau_\Delta + \tau_\Sigma) \right)
\Bigr]\omega_N \notag \\
&+\Bigl[
\tau_\Delta \left( \kappa_\Delta (\eta_\Delta + \eta_\Sigma)
+\pounds_B (\tau_\Delta - \tau_\Sigma)\right)
-\kappa_\Delta \left( \pounds_B \kappa_\Sigma + \pounds_N \tau_\Delta \right)
\Bigr]\omega_B\:,\\
0~=~& \Bigl[
\kappa_\Delta \pounds_B \tau_\Delta 
- \tau_\Sigma \pounds_N \tau_\Delta
+\kappa_\Sigma \pounds_N \kappa_\Sigma
-\pounds_N \pounds_T \kappa_\Sigma \notag\\
&+\tau_\Delta \bigl(
\kappa_\Delta (\eta_\Delta-\eta_\Sigma)
+\tau_\Sigma (\eta_\Delta+\eta_\Sigma)
+\pounds_B (\kappa_\Delta-\kappa_\Sigma)
-\pounds_N (2\tau_\Delta + \tau_\Sigma)
\bigr)\Bigr]\omega_N \notag \\
&+\Bigl[
\kappa_\Delta \pounds_N \tau_\Delta 
+\tau_\Sigma \pounds_B \tau_\Delta
+\kappa_\Sigma \pounds_B \kappa_\Sigma
-\pounds_B \pounds_T \kappa_\Sigma \notag \\
&+\tau_\Delta\bigl(
-\kappa_\Delta (\eta_\Delta + \eta_\Sigma)
+\tau_\Sigma (\eta_\Delta - \eta_\Sigma)
+\pounds_N (\kappa_\Delta + \kappa_\Sigma)
-\pounds_B (2\tau_\Delta -\tau_\Sigma)
\bigr)
\Bigr]\omega_B\:,\\
0~=~& \biggl[
\Bigl(\kappa_\Sigma-\kappa_\Delta \Bigr) \tau_\Delta^3
+\Bigl(
(\eta_\Delta -\eta_\Sigma)\pounds_B \kappa_\Sigma
+(\eta_\Delta-\eta_\Sigma) \pounds_N \tau_\Delta
-2\pounds_N \pounds_B \tau_\Delta
+2\pounds_N \pounds_N \kappa_\Sigma
\Bigr) \frac{\tau_\Delta}{2}\notag \\
&+
\frac{1}{2} \Bigl(
(\pounds_B \tau_\Delta) (\pounds_B \kappa_\Sigma)
+3(\pounds_B \tau_\Delta) (\pounds_N \tau_\Delta)
-(\pounds_B \kappa_\Sigma) (\pounds_N \kappa_\Sigma)
-3(\pounds_N \tau_\Delta) (\pounds_N \kappa_\Sigma)
\Bigr)
\biggr] \omega_N \notag\\
&+\biggl[
\tau_\Delta^4-\Bigl(\tau_\Sigma+2\tau_T \Bigr) \tau_\Delta^3
+\Bigl(\eta_\Delta^2 + \eta_\Sigma^2
+\pounds_B (\eta_\Delta - \eta_\Sigma)
-\pounds_N (\eta_\Delta + \eta_\Sigma)
\Bigr) \tau_\Delta^2 \notag \\
&+
\Bigl(
(\eta_\Delta - \eta_\Sigma)\pounds_B \tau_\Delta
-(\eta_\Delta - \eta_\Sigma) \pounds_N \kappa_\Sigma
+2\pounds_N \pounds_B \kappa_\Sigma
+2\pounds_N \pounds_N \tau_\Delta
\Bigr)\frac{\tau_\Delta}{2}
-\frac{1}{2} \Bigl( (\pounds_B \kappa_\Sigma)^2 +3(\pounds_N \tau_\Delta)^2\Bigr)
\biggr] \omega_B\:,\\
0~=~& \biggl[
\tau_\Delta^4 + \Bigl( \tau_\Sigma -2\tau_T\Bigr) \tau_\Delta^3
+\Bigl( \eta_\Delta^2 + \eta_\Sigma^2 + \pounds_B (\eta_\Delta-\eta_\Sigma)
-\pounds_N (\eta_\Delta + \eta_\Sigma)
\Bigr) \tau_\Delta^2 \notag \\
&
-\Bigl(
(\eta_\Delta + \eta_\Sigma)\pounds_B \kappa_\Sigma
+(\eta_\Delta + \eta_\Sigma) \pounds_N \tau_\Delta
-2\pounds_B \pounds_B \tau_\Delta
+2\pounds_B \pounds_B \kappa_\Sigma
\Bigr) \frac{\tau_\Delta}{2}
-\frac{1}{2} \Bigl(
(\pounds_N \kappa_\Sigma)^2
+3(\pounds_B \tau_\Delta)^2
\Bigr)
\biggr] \omega_N \notag\\
&+\biggl[
-\Bigl( \kappa_\Delta + \kappa_\Sigma\Bigr) \tau_\Delta^3
+\Bigl(
(\eta_\Delta + \eta_\Sigma) \pounds_N \kappa_\Sigma
-(\eta_\Delta +\eta_\Sigma)\pounds_B \tau_\Delta
-\pounds_B \pounds_B \kappa_\Sigma
-\pounds_B \pounds_N \tau_\Delta
\Bigr) \frac{\tau_\Delta}{2} \notag \\
&
+\frac{1}{2}\Bigl(
3(\pounds_B \tau_\Delta) (\pounds_B \kappa_\Sigma)
+3(\pounds_B \tau_\Delta) (\pounds_N \tau_\Delta)
-(\pounds_B \kappa_\Sigma) (\pounds_N \kappa_\Sigma)
-(\pounds_N \tau_\Delta) (\pounds_N \kappa_\Sigma)
\Bigr)
\biggr] \omega_B\:.
\end{align}
\label{eq:4th_obst_case1}
\end{subequations}
 }
Rewriting \eqs \eqref{eq:4th_obst_case1} as
${\boldsymbol R}^{(4)}_{\mathrm{cs. 1}} {\boldsymbol \omega} = 0$,
the rank of the {\it fourth obstruction matrix of case 1} ${\boldsymbol R}^{(4)}_{\mathrm{cs. 1}}$
governs the number of KVs in a way analogous to
that of ${\boldsymbol R}^{(1)}_{\mathrm{cs. 1}}$, ${\boldsymbol R}^{(2)}_{\mathrm{cs. 1}}$ and ${\boldsymbol R}^{(3)}_{\mathrm{cs. 1}}$.

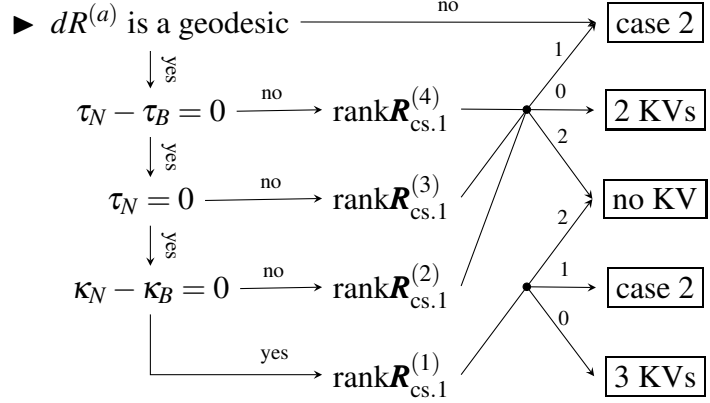
\begin{figure}[t]
		\begin{center}
			\begin{tikzpicture}[>=stealth]
			\matrix (tree) [
			matrix of nodes,
			minimum size=0.3cm,
			column sep=0.3cm,
			row sep=0.3cm,]
			{
				$\blacktriangleright \ dR^{(a)}$ is a geodesic & &\qquad&\fbox{case 2} \\
				$\tau_N - \tau_B = 0$ & $\mathrm{rank} {\boldsymbol R}_{\mathrm{cs. 1}}^{(4)}$ & & \fbox{2 KVs} \\
				$\tau_N = 0$ & $\mathrm{rank} {\boldsymbol R}_{\mathrm{cs. 1}}^{(3)}$ & & \fbox{no KV} \\
				$\kappa_N - \kappa_B=0$& $\mathrm{rank} {\boldsymbol R}_{\mathrm{cs. 1}}^{(2)}$ & & \fbox{case 2}\\
				& $\mathrm{rank} {\boldsymbol R}_{\mathrm{cs. 1}}^{(1)}$ & \quad & \fbox{3 KVs}\\
			};
			\draw[name path=INV1,white] (tree-5-3) -- (tree-1-3);
			\draw[name path=INV2,white] (tree-4-2) -- (tree-4-4);
			\draw[->] (tree-1-1) -- (tree-2-1) node [midway,above,sloped] {{\scriptsize yes}};
			\draw[->] (tree-1-1) -- (tree-1-4) node [midway,above] {{\scriptsize no}};
			\draw[->] (tree-2-1) -- (tree-2-2) node [midway,above] {\hspace{-0.2cm}{\scriptsize no}};
			\draw[->] (tree-2-1) -- (tree-3-1) node [midway,above,sloped] {{\scriptsize yes}};
			\draw[->] (tree-3-1) -- (tree-3-2) node [midway,above] {\hspace{0.2cm}{\scriptsize no}};
			\draw[->] (tree-3-1) -- (tree-4-1) node [midway,above,sloped] {{\scriptsize yes}};
			\draw[->] (tree-4-1) -- (tree-4-2) node [midway,above] {\hspace{-0.2cm}{\scriptsize no}};
			\draw[->] (tree-4-1) |- (tree-5-2) node [midway,above] {\hspace{3.2cm} {\scriptsize yes}};
			\draw[name path=to2KVs, ->] (tree-2-2) -- (tree-2-4.west) node [above, pos=0.75] {{\scriptsize $0$}};
			\path[name intersections={of=INV1 and to2KVs,by={P}}];
			\path[name intersections={of=INV1 and INV2,by={Q}}];
			\node[scale=0.25, circle, fill,draw] at (P) {};
			\node[scale=0.25, circle, fill,draw] at (Q) {};
			\draw[] (tree-3-2.east) -- (P);
			\draw[] (tree-4-2.east) -- (P);
			\draw[->] (P) -- (tree-1-4.west) node [left,pos=0.7] {{\scriptsize $1$}};
			\draw[->] (P) -- (tree-3-4.west) node [right,pos=0.3] {{\scriptsize $2$}};;
			\draw[] (tree-5-2.east) -- (Q);
			\draw[->] (Q) -- (tree-3-4.west) node [left,pos=0.8] {{\scriptsize $2$}};
			\draw[->] (Q) -- (tree-4-4.west) node [above,pos=0.55] {{\scriptsize $1$}};
			\draw[->] (Q) -- (tree-5-4.west) node [above,pos=0.55] {{\scriptsize $0$}};
			
			\end{tikzpicture}
			\caption{Our sub-algorithm for case 1, see \eqs \eqref{eq:ricci_coes_case1}, \eqref{eq:1st_obst_case1}, \eqref{eq:2nd_obst_case1},
				\eqref{eq:3rd_obst_case1} and \eqref{eq:4th_obst_case1} for notations.
				Remark that if the non-zero 1-form $dR^{(a)}$ is not a geodesic, we return to the analysis of case 2.
			}
			\label{fig:case1}
		\end{center}
	\end{figure}

We are at the end of this branch and summarise our results in Figure \ref{fig:case1}.

\section{Analysis of case 0}\label{sec:case0}
In this case, we have $dR^{(1)}=dR^{(2)}=dR^{(3)}=0$.
This implies that ${\boldsymbol R}_a$ is a zero matrix, and
all eigenvalues of the Ricci tensor are constant.
Here we take advantage of this property and
use the eigensystem of the Ricci tensor, $(E_1^a,E_2^a,E_3^a)$, as an orthonormal basis.
As we will see below, our analysis depends on {\it Segre types} of the Ricci tensor.

Since an arbitrary vector can be the annihilator of ${\boldsymbol R}_a$,
any KV can be written as
\begin{align}
K^a =  \pi_1\:E^a_1 +  \pi_2\:E^a_2 + \pi_3\:E^a_3\:,
\label{eq:KV_ansatz_case0}
\end{align}
where $(\pi_1,\pi_2,\pi_3)$ are unknown functions.
We also introduce the Ricci rotation coefficients for the eigensystem as
\begin{subequations}
\begin{align}
E_1^b \nabla_b
\begin{pmatrix}
E_1^a \\
E_2^a \\
E_3^a
\end{pmatrix}
&~=~
\begin{pmatrix}
0 	& \kappa_1 & \eta_1 \\
-\kappa_1	& 0 & \tau_1 \\
-\eta_1	& -\tau_1 & 0
\end{pmatrix}
\begin{pmatrix}
E_1^a \\
E_2^a \\
E_3^a
\end{pmatrix}\:,\\
E_2^b \nabla_b
\begin{pmatrix}
E_1^a \\
E_2^a \\
E_3^a
\end{pmatrix}
&~=~
\begin{pmatrix}
0& -\kappa_2 & \tau_2 \\
\kappa_2	& 0 & \eta_2 \\
-\tau_2	& -\eta_2 & 0
\end{pmatrix}
\begin{pmatrix}
E_1^a \\
E_2^a \\
E_3^a
\end{pmatrix}\:, \\
E_3^b \nabla_b
\begin{pmatrix}
E_1^a \\
E_2^a \\
E_3^a
\end{pmatrix}
&~=~
\begin{pmatrix}
0& \tau_3 & -\kappa_3 \\
-\tau_3 & 0 & -\eta_3 \\
\kappa_3& \eta_3 & 0
\end{pmatrix}
\begin{pmatrix}
E_1^a \\
E_2^a \\
E_3^a
\end{pmatrix}\:,
\end{align}
\end{subequations}
where
\begin{align}
&\kappa_1 ~\equiv~ E_2^a E_1^b \nabla_b (E_1)_a\:,
&&\eta_1 ~\equiv~ E_3^a E_1^b \nabla_b (E_1)_a\:,
&&\tau_1 ~\equiv~ E_3^a E_1^b \nabla_b (E_2)_a\:,\notag\\
&\kappa_2 ~\equiv~ E_1^a E_2^b \nabla_b (E_2)_a\:,
&&\eta_2 ~\equiv~ E_3^a E_2^b \nabla_b (E_2)_a\:,
&&\tau_2 ~\equiv~ E_3^a E_2^b \nabla_b (E_1)_a\:,\notag\\
&\kappa_3 ~\equiv~ E_1^a E_3^b \nabla_b (E_3)_a\:,
&&\eta_3 ~\equiv~ E_2^a E_3^b \nabla_b (E_3)_a\:,
&&\tau_3 ~\equiv~ E_2^a E_3^b \nabla_b (E_1)_a\:.
\label{eq:ricci_coes_case0}
\end{align}

Using \eqs \eqref{eq:KV_ansatz_case0} and \eqref{eq:ricci_coes_case0},
we can write out the equations of evolution of $(\pi_1,\pi_2,\pi_3)$.
However, the equation will not be used and is somewhat lengthy.
So we leave this to \ref{app:eqs_case0} and
just cite its integrability condition
\begin{subequations}
	\begin{align}
	0~=~&
	(\lambda_1-\lambda_2) \Bigl[ \kappa_2 \pi_2
		-(\tau_1 + \tau_3) \pi_3+\bar{\pi}_2
		\Bigr]\:,\\
	0~=~&(\lambda_2 - \lambda_3)
	\Bigl[(\tau_1-\tau_2) \pi_1 - \eta_3 \pi_3 - \bar{\pi}_3\Bigr]\:,\\
	0~=~&
	(\lambda_3 - \lambda_1) \Bigl[\eta_1 \pi_1 + (\tau_2-\tau_3) \pi_2  +\bar{\pi}_1\Bigr]\:,
	\end{align}
	\label{eq:int_case0}
\end{subequations}
where $\lambda_i$ are the eigenvalues of the Ricci tensor,
\begin{align}
&R^a{}_b E_i^b ~=~ \lambda_i E_i^a\:,
&& ( i ~=~ 1,2,3)
\label{eq:eigenvalues_case0}
\end{align}
and $\bar{\pi}_i$ are new variables defined as
\begin{align}
&\bar{\pi}_1 ~\equiv~ \pounds_3 \pi_1\:,
&&\bar{\pi}_2 ~\equiv~ \pounds_1 \pi_2\:,
&&\bar{\pi}_3 ~\equiv~ \pounds_2 \pi_3\:.
\label{eq:high_pi_case0}
\end{align}
Here $\pounds_i$ denotes the Lie derivative along $E_i^a$.

The integrability condition \eqref{eq:int_case0} is
trivially satisfied if the Segre type is $\left[ (111) \right]$, $\lambda_1 = \lambda_2 = \lambda_3$.
In the remaining parts of this subsection,
we discuss the Segre types $\left[ (11)1 \right]$ and $\left[ 111 \right]$.

\subsection{Branch where the Segre type is $\left[ (11)1 \right]$}
In this branch, we can assume $\lambda_1 \neq \lambda_2$ and $\lambda_2=\lambda_3$ without loss of generality.
From the constancy of the eigenvalues \eqref{eq:constancy_case0}, we have
\begin{align}
&\kappa_2 + \kappa_3~=~0\:,
&&\kappa_1~=~0\:,
&&\eta_1~=~0\:.
\label{eq:assume_case0_segre2}
\end{align}
Under \eq \eqref{eq:assume_case0_segre2},
the integrabiliy condition \eqref{eq:int_case0} reads
\begin{align}
&\bar{\pi}_1 ~=~ -(\tau_2-\tau_3)\:\pi_2\:,
&&\bar{\pi}_2 ~=~ -\kappa_2\:\pi_2 + (\tau_1 + \tau_3)\:\pi_3\:,
\label{eq:pis_case0_segre2}
\end{align}
Substituting \eq \eqref{eq:pis_case0_segre2} into \eq \eqref{eq:Killing_case0},
we obtain the equations for $(\pi_1, \pi_2, \pi_3, \bar{\pi}_3)$.
We leave its explicit form to \eq \eqref{eq:Killing_case0_segre2} in \ref{app:eqs_case0}.
The integrability condition of \eq \eqref{eq:Killing_case0_segre2} reads
\begin{subequations}
	\begin{align}
	0~=~&(\tau_2+\tau_3) \bar{\pi}_3 + \tau_2 (\tau_2 + \tau_3) \pi_1
	-(\pounds_2 \kappa_2) \pi_2 - (2\eta_2 \kappa_2 + \pounds_2 \tau_3) \pi_3\:,
	\label{eq:int_tau_case0_segre2}\\
	0~=~&2\kappa_2 \bar{\pi}_3 + 2\kappa_2 \tau_2 \pi_1 + (\pounds_2 \tau_2) \pi_2
	+(2 \eta_3 \kappa_2 + \pounds_3 \tau_2) \pi_3\:,
	\label{eq:int_k_case0_segre2}
	\\
	0~=~&[ \pounds_2(\tau_2 - \tau_3) ]\pi_2+ [ \pounds_3(\tau_2 - \tau_3)] \pi_3\:,\\
	0~=~&\left[ \frac{\tau_2 -\tau_3}{2} \pounds_2 \tau_2 - \tau_2 \pounds_2 \tau_3 
	-\kappa_2 (2\eta_3 \kappa_2+\pounds_3 \tau_2 - \eta_2 (\tau_2 + \tau_3) ) \right] \pi_2 \notag \\
	&+\left[ \frac{\tau_2 -\tau_3}{2}\pounds_3 \tau_2 - \tau_2 \pounds_3 \tau_3
	- \kappa_2 (2\eta_2 \kappa_2-\pounds_2 \tau_3 +\eta_3 (\tau_2+\tau_3))   \right] \pi_3\:.
	\end{align}
	\label{eq:int_case0_segre2}
\end{subequations}
If there are four linearly independent KVs, then \eqs \eqref{eq:int_case0_segre2} implies $\tau_2+\tau_3 = \kappa_2 = 0$.
We examine the sub-branch depending on whether $\tau_2+\tau_3$ and $\kappa_2$ are zero, or not.

\subsubsection{Sub-branch where $\tau_2+\tau_3 = \kappa_2 = 0$}
In this sub-branch, the integrability condition \eqref{eq:int_case0_segre2} gives a relation
\begin{align}
0~=~ (\pounds_2 \tau_2) \pi_2 + (\pounds_3 \tau_3) \pi_3\:.
\end{align}
On the other hand, the derivatives of the Ricci rotation coefficients \eqref{eq:ricci_coes_case0} are not independent as it is for case 1.
We collect their relations in \ref{app:curv_rels}.
From \eqs \eqref{eq:curv_rels_case0}, it follows that $\pounds_2 \tau_2=\pounds_3 \tau_3=0$ automatically.
As there are no extra conditions, we conclude that four KVs exist in this sub-branch.

\subsubsection{Sub-branch where $\tau_2+\tau_3 = 0$ but $\kappa_2 \neq 0$}
In this sub-branch, we can solve \eq \eqref{eq:int_case0_segre2} for $\bar{\pi}_3$ and obtain
\begin{align}
\bar{\pi}_3 ~=~ -\tau_2 \pi_1 - \eta_3 \pi_3\:.
\label{eq:barpi_case0_sub2}
\end{align}
By using \eqref{eq:barpi_case0_sub2}, we rewrite \eqs \eqref{eq:Killing_case0_segre2} as
\begin{align}
&\nabla_a {\boldsymbol \pi} ~=~ {\boldsymbol \Pi}^{(1)}_a {\boldsymbol \pi}\:,
&&{\boldsymbol \pi}~\equiv~
\begin{pmatrix}
\pi_1\\
\pi_2\\
\pi_3
\end{pmatrix}\:,
\end{align}
where
\begin{align}
{\boldsymbol \Pi}^{(1)}_a~\equiv~&
(E_1)_a
\begin{pmatrix}
0 & 0 & 0\\
0 & -\kappa_2 & \tau_1-\tau_2\\
0 & \tau_2-\tau_1 & \kappa_2
\end{pmatrix}
+(E_2)_a
\begin{pmatrix}
0 & 0 & 2\tau_2 \\
\kappa_2 & 0 & \eta_2 \\
-\tau_2 & 0 & -\eta_3
\end{pmatrix}
+(E_3)_a
\begin{pmatrix}
0 & -2\tau_2 & 0\\
\tau_2 & -\eta_2 & 0\\
-\kappa_2 & \eta_3 & 0
\end{pmatrix}\:.
\end{align}
Its integrability condition, $(\nabla_{[a} {\boldsymbol \Pi}^{(1)}_{b]} -
{\boldsymbol \Pi}^{(1)}_{[a} {\boldsymbol \Pi}^{(1)}_{b]}){\boldsymbol \pi} = 0$,
can be written by
\begin{align}
{\boldsymbol R}^{(1)}_{\mathrm{cs. 0}} {\boldsymbol \pi} ~=~ 0\:,
\end{align}
where
\begin{align}
{\boldsymbol R}^{(1)}_{\mathrm{cs. 0}}
~\equiv~
\begin{pmatrix}
0 & \pounds_2 \tau_2 & \pounds_3 \tau_2\\
0 & \pounds_2 \kappa_2 & \pounds_3 \kappa_2\\
\pounds_1 \eta_2 & \pounds_2 \eta_2 & \pounds_3 \eta_2\\
\pounds_1 \eta_3 & \pounds_3 \eta_3 & \pounds_3 \eta_3
\end{pmatrix}
\:,
\label{eq:1st_obst_case0}
\end{align}
is the {\it first obstruction matrix of case 0}
whose rank governs the number of KVs.
If $\mathrm{rank} {\boldsymbol R}^{(1)}_{\mathrm{cs. 0}} = 0$, three KVs exist.
If $\mathrm{rank} {\boldsymbol R}^{(1)}_{\mathrm{cs. 0}}= 3$, there is no KV.
Otherwise, when $\mathrm{rank} {\boldsymbol R}^{(1)}_{\mathrm{cs. 1}} = 1$ or $2$,
our analysis reduces to that of case 1 or 2 with appropriate identifications.
\footnote{
For instance, if $\pounds_1 \eta_2 \neq 0$, we can write $\pi_1$ as
$\pi_1 = - \frac{\pounds_2 \eta_2}{\pounds_1 \eta_2} \pi_2 - \frac{\pounds_3 \eta_2}{\pounds_1 \eta_2} \pi_3$,
leading to the form of a KV,
$K^a = \pi_2 \left( E_2^a - \frac{\pounds_2 \eta_2}{\pounds_1 \eta_2} E_1^a \right)
+\pi_3 \left( E_3^a - \frac{\pounds_3 \eta_2}{\pounds_1 \eta_2} E_1^a \right)$.
Thus our analysis goes back to that of case 1.}

\subsubsection{Sub-branch where $\tau_2+\tau_3 \neq 0$}
As similar to the previous sub-branch, \eq \eqref{eq:int_tau_case0_segre2} allows us to
put $\bar{\pi}_3$ into $(\pi_1, \pi_2, \pi_3)$
\begin{align}
\bar{\pi}_3 ~=~ -\tau_2 \pi_1 + \frac{\pounds_2 \kappa_2}{\tau_2+\tau_3}\pi_2
+\left(\frac{\pounds_3 \kappa_2}{\tau_2 + \tau_3} - \eta_3\right) \pi_3\:.
\label{eq:barpi_case0_sub3}
\end{align}
By combining \eqs \eqref{eq:barpi_case0_sub3} and \eqref{eq:Killing_case0_segre2}, we obtain
\begin{align}
&\nabla_a {\boldsymbol \pi} ~=~ {\boldsymbol \Pi}^{(2)}_a {\boldsymbol \pi}\:,
&&{\boldsymbol \pi}~\equiv~
\begin{pmatrix}
\pi_1\\
\pi_2\\
\pi_3
\end{pmatrix}\:,
\end{align}
where
\begin{align}
{\boldsymbol \Pi}^{(2)}_a~\equiv~&
(E_1)_a
\begin{pmatrix}
0 & 0 & 0\\
0 & -\kappa_2 & \tau_1+\tau_3\\
0 & \tau_2-\tau_1 & \kappa_2
\end{pmatrix}
+(E_2)_a
\begin{pmatrix}
0 & 0 & \tau_2-\tau_3 \\
\kappa_2 & 0 & \eta_2 \\
-\tau_2 & -\eta_2 + \tfrac{2\eta_3 \kappa_2 + \pounds_3 \tau_2}{\tau_2+\tau_3} & \tfrac{2\eta_2\kappa_2-\pounds_2 \tau_3}{\tau_2+\tau_3}
\end{pmatrix} \notag\\
&+(E_3)_a
\begin{pmatrix}
0 & \tau_3-\tau_2 & 0\\
-\tau_3 & -\tfrac{2\eta_3\kappa_2 + \pounds_3 \tau_2}{\tau_2+\tau_3} & -\eta_3-\tfrac{2\eta_2\kappa_2 - \pounds_2 \tau_3}{\tau_2+\tau_3}\\
-\kappa_2 & \eta_3 & 0
\end{pmatrix}\:.
\end{align}
By using the shorthand notation for the Ricci rotation coefficients,
\begin{align}
\tau_\pm ~\equiv~ \tau_3 \pm \tau_2\:,
\end{align}
its integrability condition, $(\nabla_{[a} {\boldsymbol \Pi}^{(2)}_{b]} -
{\boldsymbol \Pi}^{(2)}_{[a} {\boldsymbol \Pi}^{(2)}_{b]}){\boldsymbol \pi} = 0$,
reads
\begin{subequations}
\begin{align}
0~=~&(\pounds_2 \tau_-) \pi_2 + (\pounds_3 \tau_-) \pi_3\:,\\
0~=~&\left( 4 \pounds_2 \kappa_2^2 + \pounds_2 \tau_+^2\right) \pi_2
+ \left( 4 \pounds_3 \kappa_2^2 + \pounds_3 \tau_+^2\right) \pi_3\:,
\end{align}
and
  {\footnotesize
\begin{align}
&0~=~ \frac{1}{2}\biggl[
2\eta_2\kappa_2\tau_+
+\eta_3 \tau_+ (\tau_+ -\tau_-)
+\pounds_2 \kappa_2^2
+(\tau_- - \tau_+)\pounds_3\kappa_2
\biggr] \pi_1 \notag \\
&+\biggl[
\pounds_2 \pounds_2 \kappa_2 -2\eta_3\pounds_2 \kappa_2 + \tau_+\pounds_2 \eta_2
+\frac{(\pounds_2 \kappa_2) \pounds_2 (\tau_- - \tau_+)}{2\tau_+}
-\frac{\eta_2 \pounds_2 \kappa_2^2}{\tau_+}
+\frac{2(\pounds_2\kappa_2) (\pounds_3 \kappa_2)}{\tau_+}
\biggr] \pi_2 \notag \\
&+\biggl[
\pounds_2 \pounds_3 \kappa_2
-3 \eta_3\pounds_3 \kappa_2
+\tau_1 \tau_-\tau_+
+\eta_2 \pounds_2 \kappa_2
+\tau_+ \pounds_3 \eta_2
+ \frac{(\pounds_3\kappa_2) \pounds_2(\tau_- - \tau_+)}{2\tau_+}
-\frac{\eta_2 \pounds_3 \kappa_2^2}{\tau_+}
+\frac{2(\pounds_3 \kappa_2)^2}{\tau_+}
\biggr] \pi_3\:,\\
&0~=~ \frac{1}{2}\biggl[
2\eta_3\kappa_2\tau_+
-\eta_2 \tau_+ (\tau_+ +\tau_-)
-\pounds_3 \kappa_2^2
-(\tau_- + \tau_+)\pounds_2\kappa_2
\biggr] \pi_1 \notag \\
&+\biggl[
\pounds_3 \pounds_2 \kappa_2
-3 \eta_2\pounds_2 \kappa_2
-\tau_1 \tau_-\tau_+
+\eta_3 \pounds_3 \kappa_2
-\tau_+ \pounds_2 \eta_3
- \frac{(\pounds_2\kappa_2) \pounds_3(\tau_- + \tau_+)}{2\tau_+}
+\frac{\eta_3 \pounds_2 \kappa_2^2}{\tau_+}
-\frac{2(\pounds_2 \kappa_2)^2}{\tau_+}
\biggr] \pi_2 \notag \\
&+\biggl[
\pounds_3 \pounds_3 \kappa_2 -2\eta_2\pounds_3 \kappa_2 - \tau_+\pounds_3 \eta_3
-\frac{(\pounds_3 \kappa_2) \pounds_3 (\tau_- + \tau_+)}{2\tau_+}
+\frac{\eta_3 \pounds_3 \kappa_2^2}{\tau_+}
-\frac{2(\pounds_2\kappa_2) (\pounds_3 \kappa_2)}{\tau_+}
\biggr] \pi_3\:. 
\end{align}
\label{eq:2nd_obst_case0}
 }
\end{subequations}
Rewriting \eqs \eqref{eq:2nd_obst_case0} as ${\boldsymbol R}^{(2)}_{\mathrm{cs. 0}} {\boldsymbol \pi}=0$,
$\mathrm{rank} {\boldsymbol R}^{(2)}_{\mathrm{cs. 0}}$ controls the number of KVs
in a way analogous to that of ${\boldsymbol R}^{(1)}_{\mathrm{cs. 0}}$.

\subsection{Branch where the Segre type is $\left[111 \right]$}
\begin{figure}[ht]
		\begin{center}
			\begin{tikzpicture}[>=stealth]
			\matrix (tree) [
			matrix of nodes,
			minimum size=0.3cm,
			column sep=0.3cm,
			row sep=0.3cm,]
			{
				&$\blacktriangleright \ $Segre type of $R_{ab}$ & & \qquad& \fbox{no KV} \\
				& & $\mathrm{rank} {\boldsymbol R}_{\mathrm{cs. 0}}^{(3)}$ & &\fbox{case 2}\\
				& $\tau_2 + \tau_3 = 0$ & $\mathrm{rank} {\boldsymbol R}_{\mathrm{cs. 0}}^{(2)}$ & &\fbox{case 1}\\
				\fbox{4 KVs} & $\kappa_2 = 0$ & $\mathrm{rank} {\boldsymbol R}_{\mathrm{cs. 0}}^{(1)}$ & \quad & \fbox{3 KVs}\\
			};
				\draw[name path=INV, ->,white] (tree-4-4) -- (tree-1-4);
				\draw[->] (tree-1-2) -| (tree-2-3) node[above, pos=0.25] {{\scriptsize $\left[111\right]$}};
				\draw[->] (tree-1-2) -- (tree-3-2) node[left, pos=0.5] {{\scriptsize $\left[(11)1\right]$}};
				\draw[->] (tree-3-2) -- (tree-4-2) node[midway, above, sloped] {{\scriptsize yes}};
				\draw[->] (tree-3-2) -- (tree-3-3) node[above,pos=0.44] {{\scriptsize no}};
				\draw[->] (tree-4-2) -- (tree-4-1) node[above, pos=0.45] {{\scriptsize yes}};
				\draw[->] (tree-4-2) -- (tree-4-3) node[above, pos=0.6] {{\scriptsize no}};
				\draw[name path=toCase1, ->] (tree-3-3) -- (tree-3-5.west) node[above,pos=0.77] {{\scriptsize $1$}};
				\path[name intersections={of=INV and toCase1,by={P}}];
				\node[scale=0.25, circle, fill,draw] at (P) {};
				\draw[] (tree-2-3.east) -- (P);
				\draw[] (tree-4-3.east) -- (P);
				\draw[->] (P) -- (tree-1-5.west) node[left, pos=0.75] {{\scriptsize $3\!\!$}};
				\draw[->] (P) -- (tree-2-5.west) node[left, pos=0.81] {{\scriptsize $2$}};
				\draw[->] (P) -- (tree-4-5.west) node[right, pos=0.32] {{\scriptsize $0$}};
			
			\end{tikzpicture}
			\caption{Our sub-algorithm for case 0, see \eqs \eqref{eq:ricci_coes_case0}, \eqref{eq:1st_obst_case0}, \eqref{eq:2nd_obst_case0}
				and \eqref{eq:3rd_obst_case0} for notations.
				Remark that this algorithm depends on Segre type of the Ricci tensor.
			}
			\label{fig:case0}
		\end{center}
		\end{figure}
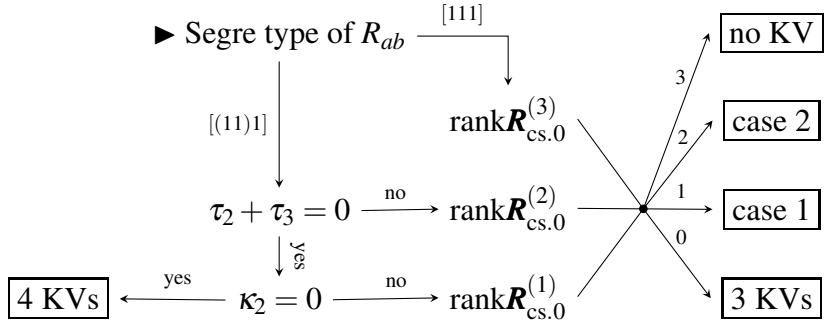

In this branch, three eigenvalues of the Ricci tensor, $(\lambda_1, \lambda_2,\lambda_3)$ differ from each other.
The constancy of the eigenvalues \eqref{eq:constancy_case0} leads to
\begin{align}
&\varphi_1 \kappa_2-\kappa_3~=~0\:,
&&\varphi_2 \eta_3 -\kappa_1~=~0\:,
&&\varphi_3 \eta_1 - \eta_2~=~0\:,
\label{eq:assume_case0_segre111}
\end{align}
where
\begin{align}
&\varphi_1 ~\equiv~ \frac{\lambda_1-\lambda_2}{\lambda_3-\lambda_1}\:,
&&\varphi_2 ~\equiv~ -\frac{\lambda_2-\lambda_3}{\lambda_1-\lambda_2}\:,
&&\varphi_3 ~\equiv~ \frac{\lambda_3-\lambda_1}{\lambda_2-\lambda_3}\:.
\end{align}
Moreover, we solve \eqs \eqref{eq:int_case0} for $(\bar{\pi}_1,\bar{\pi}_2,\bar{\pi}_3)$ and obtain
\begin{align}
&\bar{\pi}_1 ~=~ -\eta_1 \pi_1 - (\tau_2 - \tau_3) \pi_2\:,
&&\bar{\pi}_2 ~=~ -\kappa_2 \pi_2 + (\tau_1 + \tau_3) \pi_3\:,
&&\bar{\pi}_3 ~=~ (\tau_1 - \tau_2)\pi_2 -\eta_3 \pi_3\:.
\label{eq:high_pis_segre111}
\end{align}
By using \eqs \eqref{eq:assume_case0_segre111} and \eqref{eq:high_pis_segre111}, we rewrite \eqs \eqref{eq:Killing_case0} as
\begin{align}
&\nabla_a
{\boldsymbol \pi}
~=~
{\boldsymbol \Pi}_a^{(3)}
{\boldsymbol \pi}
\:,
&&
{\boldsymbol \pi}~\equiv~
\begin{pmatrix}
\pi_1\\
\pi_2\\
\pi_3
\end{pmatrix}\:,
\label{eq:Killing_case0_sub2}
\end{align}
where
\begin{align}
{\boldsymbol \Pi}_a^{(3)}
~\equiv~&
(E_1)_a
\begin{pmatrix}
0 & \varphi_2 \eta_3 & \eta_1 \\
0 & -\kappa_2 & \tau_1 + \tau_3 \\
- & \tau_2 - \tau_1 & - \varphi_1 \kappa_2
\end{pmatrix}
+(E_2)_a
\begin{pmatrix}
-\varphi_2 \eta_3 & 0 & \tau_2 - \tau_3 \\
\kappa_2 & 0 & \varphi_3 \eta_1 \\
\tau_1 - \tau_2 & 0 & -\eta_3
\end{pmatrix} \notag \\
&+(E_3)_a
\begin{pmatrix}
-\eta_1 & \tau_3 - \tau_2 & 0\\
-\tau_1 - \tau_3 & -\varphi_3 \eta_1 & 0 \\
\varphi_1 \kappa_2 & \eta_3 & 0
\end{pmatrix}
\:.
\end{align}
Its integrability condition,
$( \nabla_{[a} {\boldsymbol \Pi}^{(3)}_{b]}- {\boldsymbol \Pi}^{(3)}_{[a} {\boldsymbol \Pi}^{(3)}_{b]}){\boldsymbol \pi} = 0$, is typified by
${\boldsymbol R}^{(3)}_{\mathrm{cs. 0}} {\boldsymbol \pi} = 0$
where
\begin{align}
{\boldsymbol R}^{(3)}_{\mathrm{cs. 0}} ~\equiv~
\begin{pmatrix}
\pounds_1 \kappa_2 & \pounds_2 \kappa_2 & \pounds_3 \kappa_2 \\
\pounds_1 \eta_1 & \pounds_2 \eta_1 & \pounds_3 \eta_1 \\
\pounds_1 \eta_3 & \pounds_2 \eta_3 & \pounds_3 \eta_3 \\
\pounds_1 \tau_1 & \pounds_2 \tau_1 & \pounds_3 \tau_1 \\
\pounds_1 \tau_2 & \pounds_2 \tau_2 & \pounds_3 \tau_2 \\
\pounds_1 \tau_3 & \pounds_2 \tau_3 & \pounds_3 \tau_3
\end{pmatrix}
\:.
\label{eq:3rd_obst_case0}
\end{align}
We therefore are at the conclusion that
the rank of the {\it third obstruction matrix of case 0} controls the number of KVs
in a way analogous to that of ${\boldsymbol R}^{(1)}_{\mathrm{cs. 0}}$ and ${\boldsymbol R}^{(2)}_{\mathrm{cs. 0}}$.
We summarise our results in Figure \ref{fig:case0}.

\section{Examples}\label{sec:exam}
This section is devoted to some relatively simple applications of our algorithm.

\subsection{Hamiltonian system}\label{ss:Ham}
It is well known that
a 2-dimensional natural Hamiltonian
\begin{align}
\bar{H}(p,q)~=~\tfrac{1}{2} (p_1^2+p_2^2) + V(q_1,q_2)\:,
\end{align}
can be {\it lifted} to the 3-dimensional geodesic Hamiltonian
\begin{align}
&H(p,q)~=~\tfrac{1}{2}g^{ab}_E (q) p_a p_b\:,
&&\text{where}
&&g^{ab}_E(q) ~=~ \begin{pmatrix}
1 & 0 & 0 \\
0 & 1 & 0 \\
0 & 0 & 2V(q_1,q_2)
\end{pmatrix}\:.
\end{align}
The metric $(g_E)_{ab}$ is called the Eisenhart metric.
We examine this metric with
\begin{align}
V(q_1, q_2) ~=~ \tfrac{1}{4}(q_1^4+q_2^4) + \tfrac{\epsilon}{2} q_1^2q_2^2\:,
\end{align}
by our branching algorithm.

After simple algebra, we can see that
the first test $R_{ab} \propto g_{ab}$ fails for any $\epsilon$.
Since the geodesic flow of $(g_E)_{ab}$ has the KV $\partial_{q_3}$,
the metric passes the test $dR^{(1)}\wedge dR^{(2)} \wedge dR^{(3)} = 0$.
Moreover, we find the fact that the value of $dR^{(a)}\wedge dR^{(b)}$ depends on the value of $\epsilon$:
If $\epsilon \neq 1$, the metric arrives at the analysis of case 1, and it
passes the last two tests $\kappa_{ab}=\nabla_{[a}\Omega_{b]}=0$.
So there is only one KV in this case.

In case $\epsilon=1$, the metric passes to the analysis of case 1, and
$\nabla_a R^{(1)}$ is to be a geodesic as well as an eigenvector of $R^a{}_b$ with 
the eigenvalue $-6/(q_1^2+q_2^2)$.
The metric has the properties $\tau_N=\tau_B$, $\tau_N=0$ and $\kappa_N \neq \kappa_B$.
We see that $\mathrm{rank}{\boldsymbol R}^{(2)}_{\mathrm{cs. 1}} = 2$, and therefore there are two 
KVs: $\partial_{q_1}-\partial_{q_2},\partial_{q_3}$.
This result was known in the literature.

\subsection{Foliation of the Zipoy--Voorhees spacetime}\label{ss:ZV}
We consider the Zipoy-Voorhees spacetime ($\delta \ge 0$) \cite{Zip,V}
\begin{align}
g_\delta ~=~&\left( \frac{x+1}{x-1} \right) ^\delta
\Biggl[
(x^2-y^2)\left(\frac{x^2-1}{x^2-y^2}\right)^{\delta^2}
\left(\frac{dx^2}{x^2-1}+\frac{dy^2}{1-y^2}\right) \notag\\
&+(x^2-1)(1-y^2)dz^2\Biggr]-
\left(\frac{x-1}{x+1}\right)^\delta\,dt^2\:.
\end{align}
This has Lorentzian signature in the region $|y|<1<|x|$, and $\partial_t$ is a timelike KV with
orthogonal Frobenius-integrable distribution.
Restriction of the metric on a leaf of the corresponding foliation is Riemannian:
\begin{align}
\bar{g}_\delta~=~g_\delta|_{t=\text{const}}.
\end{align}
The original $g_\delta$ is Ricci-flat, but $\bar{g}_\delta$ is flat only for $\delta=0$. We get:
$\text{rank}{\boldsymbol R}_a=0$ for $\delta=0$,
$\text{rank}{\boldsymbol R}_a=1$ for $\delta=1$,
and $\text{rank}{\boldsymbol R}_a=2$ else.

Our algorithm implies the number $\bar{d}$ of KVs for $\bar{g}_\delta$: 
6 KVs for $\delta=0$ (this corresponds to the Minkowski metric),
3 KVs for $\delta=1$ (this corresponds to the Schwarzschild metric)
and 1 KV for $\delta\neq0,1$. The number $d$ of KVs for $g_\delta$ is equal to $\bar{d}+1$.

Further computations yield that $g_\delta$ has no KT of order 2 that are not combinations of 
symmetric products of known KVs, cf.\ \cite{BV}.

\section{Conclusion}\label{sec:conc}

In this paper we demonstrated an algorithm computing dimension $d$ of the local group of motions 
of a 3-dimensional Riemannian manifold $(M,g_{ab})$ via differential invariants of $g_{ab}$.
The following properties are known.
The maximal value of $d$ is 6, which is realised if and only if $g_{ab}$ is an Einstein metric: $R^a{}_b=\frac13R\delta^a{}_b$.
This is equivalently characterised by the equations $R^{(1)}=R$, $R^{(2)}=\frac13R^2$, $R^{(3)}=\frac19R^3$.
The submaximal value of $d$ is 4, and metrics $g_{ab}$ with that many KVs were characterised
by differential invariants in \cite{Kr}. Our contribution is to go beyond this value and
determine when $d$ can be 3, 2, 1 or 0.

It would be worthwhile to comment relations of our method to that of Cartan.
Cartan's equivalence method \cite{C} provides a canonical frame on the principal 
$SO(g)$-bundle over $M$, called the Cartan bundle, and this allows to determine 
the number of KVs in principle. The Cartan-Karlhede algorithm is designed, in particular, 
for such purposes. However, to our knowledge, this was not used to solve the problem
we address in this paper. In addition, the invariants the algorithm produces are functions 
on the Cartan bundle,
while our invariants are defined on its base $M$. When $d\leq3$ and $r=0$ (see \eq \eqref{eq:dim_iso};
$r$ always vanishes for $d<3$) the frame fixing allows to pull down the invariants from the Cartan-Karlhede algorithm to the base, but our algorithm works universally for all cases.

The method we discuss in this paper works also for a large class of Lorentzian metrics in dimension 3, 
however there exist metrics with vanishing scalar invariants and no symmetries \cite{KM}. 
Such metrics have received an increasing attention in general relativity, see e.g.\ \cite{CMP,CHP}.
Technically, they fail our algorithm  because of normalisation, like the factor $U$ in Section \ref{sec:case2}.
It is possible to include such cases into consideration via additional branching, but we leave these elaborations to a future work.

Moreover, by taking into account the Weyl tensor, our method is applicable for higher-dimensional spaces.
For instance, in dimension 4 we get a $4 \times 10$ obstruction matrix
\begin{align}
({\boldsymbol R}_a,{\boldsymbol W}_a)~=~
\bigl(\nabla_a R^{(1)},\dots,\nabla_a R^{(4)},
\nabla_a W^{(1)},\dots,\nabla_a W^{(6)}\bigr)^T\:,
\end{align}
with $4\times4$ minors giving obstructions, playing a role similar to \eqref{boldR}.
Here $W^{(i)}$ are principal traces of the $i$-th powers of the Weyl tensor, considered as 
an endomorphism of $\Lambda^2TM$.
A criterion on existence of KVs in 4D should be based on this obstruction matrix.

The algorithm we designed can be applied to determine exact number of KVs for 4-dimensional 
spacetimes possessing a timelike KV with integrable orthogonal distribution, as in the case of 
Zipoy--Voorhees metric considered in Section \ref{ss:ZV}. In such situation the 
quotient along trajectories of the given timelike KV determines a local submersion of the 
Lorentzian 4-manifold onto a Riemannian 3-manifold, and our analysis of KVs can be invoked to 
simplify computation of additional isometries. Further consideration of this shall be done elsewhere.

\section*{Acknowledgment}

KT thanks Tsuyoshi Houri and Yukinori Yasui for discussions on the early stage of this work,
and Masato Nozawa for his useful comments.
Both authors are indebted to Vladimir Matveev for his advices and support;
KT is also grateful to the University of Jena for hospitality.

\appendix

\section{Relations between the Ricci rotation coefficients and the commutation relations}\label{app:curv_rels}
Generally, the derivatives of the Ricci rotation coefficients are not independent.
In this Appendix, we record the relations among \eqs \eqref{eq:ricci_coes_case1} and \eqref{eq:ricci_coes_case0}.
We also write the commutation relations of the orthonormal basis used in Sections \ref{sec:case1} and \ref{sec:case0}.

\subsection{For analysis of case 1}\label{app:curv_rels_case1}
When $\nabla_a R$ is a geodesic, that is $\kappa_T=\eta_T=0$, the following relations hold true.

\noindent
{\bf The components of the Ricci tensor:}
\begin{subequations}
 \begin{align}
\hspace{-1cm}	R_{ab}T^aT^b~=~& \pounds_T (\kappa_N+\kappa_B)- \kappa_N^2- \kappa_B^2 - 2\tau_B \tau_N\:,\\
\hspace{-1cm}	R_{ab}N^aN^b~=~& \pounds_B \eta_N + \pounds_T\kappa_T + \pounds_N \eta_B- \kappa_N (\kappa_N+ \kappa_B) - \eta_N^2 - \eta_B^2
	+2\tau_T \tau_B\:,\\
\hspace{-1cm}	R_{ab}B^aB^b~=~& \pounds_B \eta_N + \pounds_T\kappa_B + \pounds_N \eta_B- \kappa_B (\kappa_N+ \kappa_B) - \eta_N^2 - \eta_B^2
	-2\tau_T \tau_N\:,\\
\hspace{-1cm}	R_{ab}T^aN^b~=~& \pounds_B \tau_N +\pounds_N \kappa_B+\eta_B (\kappa_N-\kappa_B) - \eta_N (\tau_N + \tau_B)\:,\\
	~=~& \pounds_B \tau_T +\pounds_T \eta_B- \kappa_B \eta_B- \eta_N (\tau_T+\tau_B) \:,\\
\hspace{-1cm}	R_{ab}T^aB^b~=~& \pounds_B \kappa_N +\pounds_N \tau_B -\eta_N (\kappa_N-\kappa_B) - \eta_B (\tau_N + \tau_B) \:,\\
	~=~&\pounds_T \eta_N - \pounds_N \tau_T-\eta_N \kappa_N + \eta_B (\tau_T-\tau_N) \:,\\
\hspace{-1cm}	R_{ab}N^aB^b~=~& - \pounds_T \tau_B +\tau_B (\kappa_N + \kappa_B) +\tau_T( \kappa_N - \kappa_B )\:,\\
	~=~&-\pounds_T \tau_N+\tau_T (\kappa_N - \kappa_B) +\tau_N (\kappa_N + \kappa_B)\:.
 \end{align}
\label{eq:curv_rels_case1}
\end{subequations}

\noindent
{\bf The commutation relations:}
\begin{subequations}
	\begin{align}
	\left[ T,N \right]^a &~=~ \kappa_N N^a +(\tau_T-\tau_N) B^a\:,\\
	\left[ T,B \right]^a &~=~ -(\tau_T+\tau_B) N^a + \kappa_B B^a\:,\\
	\left[ N,B \right]^a &~=~ -(\tau_N-\tau_B) T^a -\eta_N N^a + \eta_B B^a\:.
	\end{align}
\end{subequations}

\subsection{For analysis of case 0}\label{app:curv_rels_case0}
In case 0, the following relations hold true.

\noindent
{\bf The components of the Ricci tensor:}
\begin{subequations}
 \begin{align}
\hspace{-0.4cm} R_{ab}E_1^aE_1^b~=~& \pounds_1 (\kappa_2+\kappa_3) + \pounds_2 \kappa_1 +\pounds_3 \eta_1- \kappa_1 (\kappa_1 + \eta_3) - \kappa_2^2 - \kappa_3^2 - \eta_1 \eta_2
		-2\tau_2 \tau_3\:,
		\label{eq:curv_rels_11_case0}\\
\hspace{-0.4cm} R_{ab}E_2^aE_2^b~=~& \pounds_1 \kappa_2 + \pounds_2(\kappa_1+\eta_3) +\pounds_3\eta_2- \kappa_1^2 - \kappa_2 (\kappa_2 + \kappa_3) - \eta_2(\eta_1+\eta_2)
		-\eta_3^2 +2\tau_1 \tau_3\:,
		\label{eq:curv_rels_22_case0}\\
\hspace{-0.4cm} R_{ab}E_3^aE_3^b~=~&\pounds_1 \kappa_3 + \pounds_2 \eta_3 +\pounds_3(\eta_1+\eta_2)-\kappa_3 (\kappa_2+\kappa_3) - \eta_1^2 - \eta_2^2 - \eta_3 (\kappa_1 + \eta_3)
		-2\tau_1 \tau_2\:,
		\label{eq:curv_rels_33_case0}\\
\hspace{-0.4cm} R_{ab}E_1^aE_2^b~=~&
		\pounds_2\kappa_3 + \pounds_3 \tau_2+\eta_3(\kappa_2-\kappa_3) - \tau_2(\eta_1 + \eta_2) +\tau_3(\eta_1 - \eta_2)\:,\\
		~=~&
		\pounds_1\eta_3 + \pounds_3\tau_1+\kappa_3(\kappa_1 -\eta_3)-\tau_1(\eta_1+\eta_2)+\tau_3(\eta_1-\eta_2)\:,\\
\hspace{-0.4cm} R_{ab}E_1^aE_3^b~=~&
		\pounds_2\tau_3 +\pounds_3\kappa_2-\eta_2(\kappa_2-\kappa_3)-\tau_2 (\eta_3-\kappa_1)-\tau_3(\kappa_1+\eta_3) \:,\\
		~=~&
		\pounds_1 \eta_2 - \pounds_2\tau_1+\kappa_2(\eta_1-\eta_2) +\tau_1(\eta_3 +\kappa_1) + \tau_2(\kappa_1-\eta_3) \:,\\
\hspace{-0.4cm} R_{ab}E_2^aE_3^b~=~&
		\pounds_3\kappa_1-\pounds_1\tau_3-\eta_1(\kappa_1-\eta_3)+\tau_1 (\kappa_2-\kappa_3)+\tau_3(\kappa_2+\kappa_3)\:,\\
		~=~&
		\pounds_2\eta_1- \pounds_1 \tau_2-\kappa_1(\eta_1-\eta_2) +\tau_1 (\kappa_2-\kappa_3) +\tau_2(\kappa_2+\kappa_3)\:.
 \end{align}
\label{eq:curv_rels_case0}
\end{subequations}

\noindent
{\bf The commutation relations:}
\begin{subequations}
 \begin{align}
		\left[ E_1,E_2 \right]^a &~=~ -\kappa_1 E_1^a + \kappa_2 E_2^a +(\tau_1 - \tau_2) E_3^a\:,\\
		\left[ E_1,E_3 \right]^a &~=~ -\eta_1 E_1^a -(\tau_1+\tau_3) E_2^a +\kappa_3 E_3^a\:,\\
		\left[ E_2,E_3 \right]^a &~=~ -(\tau_2-\tau_3) E_1^a -\eta_2 E_2^a + \eta_3 E_3^a\:.
 \end{align}
 \end{subequations}

In case 0, all the eigenvalues of the Ricci tensor are constant. This fact leads to
the additional relations from the Lie derivatives of \eqs \eqref{eq:curv_rels_11_case0}--\eqref{eq:curv_rels_33_case0}:
\begin{align}
(\lambda_1 - \lambda_2)\kappa_2 - (\lambda_3 - \lambda_1)\kappa_3 =
(\lambda_1 - \lambda_2)\kappa_1 - (\lambda_2 - \lambda_3)\eta_3 =
(\lambda_3 - \lambda_1)\eta_1 - (\lambda_2 - \lambda_3)\eta_2 =0\:,
\label{eq:constancy_case0}
\end{align}
where $(\lambda_1, \lambda_2, \lambda_3)$ are the eigenvalues defined in \eq \eqref{eq:eigenvalues_case0}.

\section{Supplements for case 0}\label{app:eqs_case0}
In this Appendix, we make up for deficiencies in Section \ref{sec:case0}.

We firstly write the equations of evolution of $(\pi_1, \pi_2, \pi_3)$.
By using the concrete form of a KV \eqref{eq:KV_ansatz_case0},
we can write out the Killing equation as follows.
\begin{subequations}
\begin{align}
	\pounds_1 \pi_1 ~=~& \kappa_1 \pi_2 + \eta_1 \pi_3\:,\\
	\pounds_2 \pi_1 ~=~&-\kappa_1 \pi_1 -\kappa_2 \pi_2 +(\tau_1 + \tau_2)\pi_3-\bar{\pi}_2 \:,\\
	\pounds_3 \pi_1 ~=~&\bar{\pi}_1\:,\\
	\pounds_1 \pi_2 ~=~& \bar{\pi}_2\:,\\
	\pounds_2 \pi_2 ~=~&\kappa_2 \pi_1 + \eta_2 \pi_3\:,\\
	\pounds_3 \pi_2 ~=~&-(\tau_2 + \tau_3) \pi_1 -\eta_2 \pi_2 -\eta_3 \pi_3 -\bar{\pi}_3 \:,\\
	\pounds_1 \pi_3 ~=~& -\eta_1 \pi_1 -(\tau_1 - \tau_3)\pi_2 - \kappa_3 \pi_3 - \bar{\pi}_1\:,\\
	\pounds_2 \pi_3 ~=~&\bar{\pi}_3\:,\\
	\pounds_3 \pi_3 ~=~&\kappa_3 \pi_1 +\eta_3 \pi_2\:.
\end{align}
where $(\bar{\pi}_1,\bar{\pi}_2,\bar{\pi}_3)$ are defined in \eq \eqref{eq:high_pi_case0}.
The above equations are not closed with respect to
unknown scalars $(\pi_1,\pi_2,\pi_3,\bar{\pi}_1,\bar{\pi}_2,\bar{\pi}_3)$.
We need the information of the second order derivatives of $(\pi_1,\pi_2,\pi_3)$.
From the identities $\nabla_{[a} \nabla_{b]} \pi_1 =\nabla_{[a} \nabla_{b]} \pi_2 =\nabla_{[a} \nabla_{b]} \pi_3 =$,
we obtain
\begin{align}
\pounds_1 \bar{\pi}_1 ~=~& \bigl[\kappa_1 (\tau_1-\tau_2)+\kappa_3\eta_1\bigr] \pi_1
+\bigl[ \pounds_3 \kappa_1 - \eta_1 (\kappa_1-\eta_3) - \kappa_1\eta_2 +\kappa_2(\tau_1 + \tau_3)\bigr]\pi_2\notag \\
&+\bigl[ \pounds_3 \eta_1 -\eta_1^2 - \kappa_1\eta_3 - (\tau_1+\tau_3)(\tau_1 + \tau_2)\bigr]\pi_3
+\kappa_3\bar{\pi}_1 + (\tau_1+\tau_3)\bar{\pi}_2 - \kappa_1 \bar{\pi}_3\:,\\
\pounds_2 \bar{\pi}_1 ~=~& \bigl[\kappa_1\eta_2 + \kappa_2(\tau_2+\tau_3)\bigr] \pi_1
-\bigl[ \pounds_3 \kappa_2 -\eta_2(2\kappa_2-\kappa_3) - \eta_3(\tau_2+\tau_3) +\kappa_1(\tau_2-\tau_3) \bigr] \pi_2 \notag\\
&-\bigl[ \pounds_2 \kappa_3+\eta_2(\tau_1-\tau_3) \bigr] \pi_3 +\eta_2 \bar{\pi}_2 + (\kappa_2-\kappa_3)\bar{\pi}_3\:,\\
\pounds_3 \bar{\pi}_1 ~=~& -\bigl[ \pounds_1 \kappa_3 + \pounds_3 \eta_1-\eta_1^2 - (\tau_1-\tau_3)(\tau_2+\tau_3) \bigr] \pi_1
+\bigl[ \pounds_3 \tau_3-\kappa_3\eta_3 - 2\eta_2 \tau_3 \bigr] \pi_2 \notag\\
&-\bigl[ \pounds_3 \kappa_3-\eta_3(\tau_1-\tau_3)\bigr] \pi_3
-\eta_3\bar{\pi}_2-2\tau_3 \bar{\pi}_3\:,\\
\pounds_1 \bar{\pi}_2 ~=~& -\bigl[ \pounds_1 \kappa_1 +\eta_1(\tau_1+\tau_2) \bigr] \pi_1
-\bigl[ \pounds_1 \kappa_2 +\pounds_2\kappa_1-\kappa_2^2  + (\tau_1-\tau_3)(\tau_1+\tau_2) \bigr] \pi_2 \notag\\
&-\bigl[ \pounds_1\tau_1-\kappa_1\eta_1-2\kappa_3\tau_1 \bigr] \pi_3
-2\tau_1 \bar{\pi}_1 -\eta_1 \bar{\pi}_3\:,\\
\pounds_2 \bar{\pi}_2 ~=~& \bigl[ \pounds_1\kappa_2-\eta_1\eta_2-\kappa_2^2 + (\tau_2+\tau_3)(\tau_1-\tau_2)\bigr] \pi_1
+\bigl[ \kappa_1\kappa_2 -\eta_2(\tau_2-\tau_3) \bigr] \pi_2\notag\\
&+\bigl[ \pounds_1 \eta_2 -\kappa_3\eta_2+(\kappa_2 + \eta_3)(\tau_1-\tau_2) \bigr] \pi_3
-\eta_2\bar{\pi}_1 + \kappa_1\bar{\pi}_2 + (\tau_1-\tau_2)\bar{\pi}_3\:,\\
\pounds_3 \bar{\pi}_2 ~=~& -\bigl[ \pounds_3 \kappa_1 -\kappa_3(\tau_1+\tau_2)\bigr] \pi_1
+\bigl[ \kappa_3\eta_2 + \eta_3(\tau_1-\tau_3) \bigr] \pi_2\notag\\
&-\bigl[ \pounds_1 \eta_3+\kappa_3(\kappa_1-2\eta_3) - (\tau_1-\tau_3)(\eta_1+\eta_2)  \bigr] \pi_3
-(\kappa_1-\eta_3)\bar{\pi}_1 + \kappa_3\bar{\pi}_3\:,\\
\pounds_1 \bar{\pi}_3 ~=~& -\bigl[ \pounds_2\eta_1-\kappa_1(2\eta_1-\eta_2) +\tau_1(\kappa_2-\kappa_3)+\tau_2(\kappa_2+\kappa_3)\bigr] \pi_1
-\bigl[ \pounds_1\eta_2+\kappa_1(\tau_2+\tau_3)\bigr] \pi_2\notag\\
&+\bigl[ \kappa_1\kappa_3-\eta_1(\tau_1+\tau_2) \bigr] \pi_3
+\kappa_1\bar{\pi}_1 + (\eta_1-\eta_2)\bar{\pi}_2\:,\\
\pounds_2 \bar{\pi}_3 ~=~& -\bigl[ \pounds_2 \tau_2+\kappa_2\eta_2-2\kappa_1\tau_2\bigr] \pi_1
-\bigl[ \pounds_2 \eta_2-\kappa_2(\tau_2+\tau_3) \bigr] \pi_2\notag\\
&-\bigl[ \pounds_2\eta_3 +\pounds_3\eta_2-\eta_3^2 +(\tau_2+\tau_3)(\tau_1+\tau_2) \bigr] \pi_3
-\kappa_2\bar{\pi}_1 + 2\tau_2 \bar{\pi}_2\:,\\
\pounds_3 \bar{\pi}_3 ~=~& \bigl[ \pounds_2\kappa_3+\eta_3(\kappa_2-\kappa_3) -\kappa_1\kappa_3 -\eta_1(\tau_2-\tau_3)\bigr] \pi_1
+\bigl[ \pounds_2\eta_3 -(\tau_1-\tau_3)(\tau_2-\tau_3) \notag\\
&-\eta_3^2 -\kappa_2\kappa_3\bigr] \pi_2
-\bigl[ \eta_2\eta_3 + \kappa_3(\tau_1+\tau_3) \bigr] \pi_3
-(\tau_2-\tau_3)\bar{\pi}_1-\kappa_3\bar{\pi}_2 + \eta_2 \bar{\pi}_3\:.
\end{align}
\label{eq:Killing_case0}
\end{subequations}
Its integrability condition is given by \eqs \eqref{eq:int_case0}.

If the Ricci tensor has the Segre type $\left[ (11)1 \right]$,
it follows from \eqs \eqref{eq:assume_case0_segre2} and \eqref{eq:pis_case0_segre2}
that the closed system \eqref{eq:Killing_case0} takes the form
\begin{subequations}
	\begin{align}
\hspace{-0.25cm} \pounds_1 \pi_1 \,=\, & 0\:,\\
\hspace{-0.25cm} 	\pounds_2 \pi_1 \,=\, & (\tau_2-\tau_3) \pi_3\:,\\
\hspace{-0.25cm} 	\pounds_3 \pi_1\,=\, &-(\tau_2-\tau_3)\:\pi_2\:,\\
\hspace{-0.25cm} 	\pounds_1 \pi_2 \,=\, & -\kappa_2\pi_2 + (\tau_1 + \tau_3)\pi_3\:,\\
\hspace{-0.25cm} 	\pounds_2 \pi_2 \,=\, &\kappa_2 \pi_1 + \eta_2 \pi_3\:,\\
\hspace{-0.25cm} 	\pounds_3 \pi_2 \,=\, &-(\tau_2 + \tau_3) \pi_1 -\eta_2 \pi_2 -\eta_3 \pi_3 -\bar{\pi}_3 \:,\\
\hspace{-0.25cm} 	\pounds_1 \pi_3 \,=\, & - (\tau_1+\tau_2)\pi_2 + \kappa_2 \pi_3\:,\\
\hspace{-0.25cm} 	\pounds_2 \pi_3 \,=\, &\bar{\pi}_3\:,\\
\hspace{-0.25cm} 	\pounds_3 \pi_3 \,=\, &-\kappa_2 \pi_1 +\eta_3 \pi_2\:,\\
\hspace{-0.25cm} 	\pounds_1 \bar{\pi}_3 \,=\, & -2\kappa_2 \tau_1 \pi_1 - (\pounds_1 \eta_2 - \kappa_2\eta_2) \pi_2
	-\eta_2(\tau_1+\tau_3)\pi_3\:,\\
\hspace{-0.25cm} 	\pounds_2 \bar{\pi}_3 \,=\, &-(\pounds_2 \tau_2 + \kappa_2 \eta_2)\pi_1 - (\pounds_2 \eta_2)\pi_2
	-(\pounds_2 \eta_3 + \pounds_3 \eta_2 -\eta_3^2 +(\tau_3-\tau_2) (\tau_1 -\tau_2))\pi_3\:,\!\!\\
\hspace{-0.25cm} 	\pounds_3 \bar{\pi}_3 \,=\, &-(\pounds_2\kappa_2 -2\kappa_2\eta_3)\pi_1
	+(\pounds_2 \eta_3 -\eta_3^2 +(\tau_3-\tau_2) (\tau_1 - \tau_2))\pi_2
	+\eta_2\eta_3 \pi_3 + \eta_2 \bar{\pi}_3\:.\!\!
	\end{align}
\label{eq:Killing_case0_segre2}
\end{subequations}

\end{document}